\def\gtorder{\mathrel{\raise.3ex\hbox{$>$}\mkern-14mu
    \lower0.6ex\hbox{$\sim$}}}
\def\ltorder{\mathrel{\raise.3ex\hbox{$<$}\mkern-14mu
    \lower0.6ex\hbox{$\sim$}}}

\documentclass[fleqn,usenatbib]{mnras}

\usepackage{newtxtext,newtxmath}

\usepackage[T1]{fontenc}

\DeclareRobustCommand{\VAN}[3]{#2}
\let\VANthebibliography\thebibliography
\def\thebibliography{\DeclareRobustCommand{\VAN}[3]{##3}\VANthebibliography}

\usepackage{graphicx}	
\usepackage{amsmath}	
\usepackage{multirow}
\usepackage{dirtytalk}
\usepackage{cancel}
\usepackage[normalem]{ulem}
\usepackage{appendix}

\newcommand{\ZZ}[1]{}  

\title[Bars with spinning halos and bulges]{Evolution of Stellar Bars in Spinning Dark Matter Halos\\
   and Stellar Bulges}

\author[Li et al.]{
Xingchen Li,$^{1}$\thanks{E-mail: xingchen.li@uky.edu}
Isaac Shlosman,$^{1,2}$\thanks{E-mail: isaac.shlosman@uky.edu}
Daniel Pfenniger,$^{3}$
and Clayton Heller,$^{4}$
\\
$^{1}$\rm Department of Physics \& Astronomy, University of Kentucky, Lexington, KY 40506, USA\\
$^{2}$\rm Theoretical Astrophysics, School of Sciences, Osaka University, Osaka 560-0043, Japan\\
$^{3}$\rm University of Geneva, Geneva Observatory, ch. Pegasi 51, 1290 Versoix, Switzerland\\
$^{4}$\rm Department of Physics \& Astronomy, Georgia Southern University, Statesboro, GA 30460, USA
}

\date{Accepted XXX. Received YYY; in original form ZZZ}

\pubyear{2015}

\begin{document}
\label{firstpage}
\pagerange{\pageref{firstpage}--\pageref{lastpage}}
\maketitle

\begin{abstract}
We use high-resolution numerical simulations to follow the  barred disk evolution in a suite of models with progressively more massive stellar bulges, with bulge-to-total (disk$+$bulge) mass ratios of $B/T\sim 0-0.25$, embedded in dark matter (DM) halos with the spin $\uplambda\sim 0 - 0.09$. We focus on models with a sequence of initial rotational support for bulges, and analyze their spinup and spindown. We find that (1) the presence of a bulge affects evolution of stellar bars, i.e., the timescale of bar instability, bar pattern speed and its decay, and the vertical buckling instability. The bar strength is nearly independent of $B/T$ in halos with spin $\uplambda = 0$, and is suppressed by a factor $\sim 2$ for halos with $\uplambda = 0.09$; (2) The main effect of the bulge is the destruction of the harmonic core which affects the buckling; (3) The bulge plays a minor role in the exchange of angular momentum between the barred disk and the DM halo, during its spinup and spindown; (4) Most interestingly, the buckling process triggers different response above/below the disk midplane, which anti-correlates with the bulge mass; (5) In spinning halos, the buckling process has a prolonged amplitude tail, extending by few Gyr, as verified by measuring distortions in the Laplace plane; (6) Furthermore, as verified by orbital spectral analysis, the bulge gains its spin from the bar mainly via the inner Lindblad resonance, while losing it via a number of resonances lying between the outer and inner Lindblad resonance.
\end{abstract}




\section{Introduction}
\label{sec:intro}

Galaxies are commonly thought to reside in dark matter (DM) halos whose properties follow from the larger scale filamentary structure in the universe --- the cosmic web. These halos can be characterized by a number of parameters, such as mass, spin, baryon fraction, environment, and more. While halos are far from being supported by rotation, this is a net characterization, and various components within halos can still harbor a substantial amount of angular momentum which can potentially be transferred most easily towards the halo. This angular momentum appears to have significant dynamical and secular effects on the evolution of embedded stellar disks, especially when the latter ones are barred \citep[][]{collier18,collier19a,collier19b,li23b}. 

Galaxies possess an additional spheroidal component to the DM component --- the classical stellar bulges. While formation and evolution of these bulges have been extensively investigated over the last 50 years, their interaction with barred disks still has a number of unanswered questions. While work on the spinup of these bulges has been performed \citep[e.g.,][]{sell80,saha12,saha16}, the parameter space involving the range of the bulge mass fraction of the total parent galactic disk$+$bulge, $B/T$, deserves more careful study. Moreover, most the classical bulges studied so far have been nonrotating initially, and it is not clear how the angular momentum evolution of spinning bulges behaves. The focus of our analysis, therefore, is (1) to investigate the angular momentum flow in the barred galaxies with a range of $B/T$ and the initial spin of these bulges, and (2) to investigate evolution of stellar bars in spinning bulges and DM halos under various initial conditions.

Classical bulges have been studied both observationally \citep[e.g.,][]{kormendy82,korm82,kormendy04,cappell07}, theoretically \citep[e.g.,][]{eggen62,kauffmann93} and numerically \citep[e.g.,][]{sell80,saha12,saha16}. 

A large sample of local galaxies using the optical $r$-band imaging, including the early and late-type disks have shown that the bar fraction is anticorrelating with the bulge-to-total mass ratio, and rising from 40\% to 70\% \citep{barazza08}. On the other hand, \citet{hoyle11} found that bulge-dominated galaxies host longer stellar bars compared to bulgeless galaxies. This point has been also addressed by \citet{sell80} and \citet{athana80}.

Numerical simulations have demonstrated that bulges can exert a substantial influence both on the bar instability and on the vertical buckling instability in stellar bars \citep[e.g.,][]{sell20}.  \citet{kataria18} found that massive bulges prohibit the bar formation when the ratio of the radial force due to bulge to that of the galaxy at the disk scalelength radius is larger than 0.35. Large velocity dispersion weakens the bar instability in the disk with such massive bulges \citep[e.g.,][]{athana83}. Furthermore, \citet{kataria19} found that the slowdown rate of the stellar bar is stronger in the presence of a more massive bulge, possibly due to the massive bulge gaining more angular momentum from the barred disk.

Most models have employed an initially nonspinning bulge and aimed to understand the mechanism of its spinup \citep[e.g.,][]{saha12, saha16}, its effect on the stellar bar properties, e.g., vertical buckling, its vertical thickening \citep{sell20}, and specific peanut/boxy bulge shape \citep{combes81,combes90,friedli90,pfen91,raha91,bureau05,marti06,li23a}. Observations of peanut/boxy bulges in barred galaxies have been linked to bars \citep{lutticke00,erwin17} and been detected up to $z\sim 1$ \citep{kruk19}. 

It was also demonstrated that the boxy/peanut bulges can be formed by a variety of processes and may even bypass the break of the vertical symmetry in the presence of the vertical inner Lindblad resonance (vILR) \citep{friedli90}. \citet{sell20} analyzed three such mechanisms, namely, the usual vertical buckling of a bar, the heating of the bar by the vILR in the presence of a nuclear cluster and vertical heating when the vertical symmetry is enforced. Note that the nuclear cluster had a radius of $\sim 200$\,pc, which was also the gravitational softening in the simulations.

\citet{sell80} was the first to simulate the angular momentum transfer between a stellar bar and a live DM halo in numerical simulations, while \citet{wein85}  studied the interaction between a rigid disk bar with a spheroidal component. \cite{saha12} analyzed the interaction between a low bulge-to-disk mass ratio, $B/D\sim 0.067$, with a bar, with initially nonrotating bulge, and the angular momentum channeled to the bulge via resonances, especially via the 1:1 resonance (Lagrange points) and the inner Linblad resonance (ILR) 2:1. The bulge was found to spin up and to become an anisotropic triaxial object embedded in a boxy bulge, developing a near cylindrical rotation. \citet{collier19a,collier19b} has analyzed the retrograde orbits in spheroidal components and found that such orbits can absorb the retrograde angular momentum. 

The bulge has been modeled with $B/D \sim 6.7\%$ for 3\,Gyr, with the initial spin measured in rotational-to-dispersion velocities, rather than in the angular momentum units \citep{saha13}. It has affirmed that initially spinless bulge gains more angular momentum. The goal of this work was to study the boxy/peanut-shaped  bulge evolution when its classical progenitor is spinning. The bar amplitude and size have been found to be reduced by $\sim 10-40\%$. Finally, \citet{saha16} studied  a spinup of initially spinless bulges, using bulges with $B/D\sim 11-43\%$.  

Recent modeling of isolated disk galaxies have shown that parent DM halos have a two-fold effect on the stellar bar development, which depends on the halo spin $\uplambda$ and its DM density \citep{li23b}. The halo spin has been confirmed to speed up the bar instability as noted before \citep{saha13,long14,collier18}. The decrease in the DM density has led to the appearance of a plateau in the bar strength during which the bar pattern speed remains constant.
The buckling instability has been found to be separated from the maximum of the bar strength by this plateau whose length can be few Gyrs.

In this work, we study the effect of initial conditions in the bulge-disk-halo system on the stellar bar evolution. We use a sequence of bulge masses and spins in tandem with varying the parent DM halo spin.  We aim at obtaining the general picture for bar evolution in the presence of a range of bulge and DM halo properties.

This paper is structured as following. Section\,2 describes the numerical methods used, our results are presented in section\,3, which is followed by discussion and conclusions.

\section{Numerical Methods}
\label{sec:numerics}

We use the $N$-body part of the mesh-free hydrodynamics code \textsc{gizmo} \citep{hopk15}, an extension of the \textsc{gadget-2} code \citep{sprin05}. The models of stellar disk galaxies with central spherical bulges embedded in spherical DM halos have been constructed with different bulge-to-disk mass ratios and different spin in two spherical components, the stellar bulges and DM halos. 

The units of mass, length, and velocity have been chosen as $10^{10}\,\mathrm{M_{\odot}}$, $1\,\mathrm{kpc}$, and $1\, \mathrm{km\,s^{-1}}$, respectively. As a result, the unit of time is approximately $1\,\mathrm{Gyr}$. The number of DM halo particles has been taken as $N_\mathrm{DM} = 7.2 \times 10^6$, and the total stellar particles (disk plus bulge) has been fixed to $N_\mathrm{S} = 8 \times 10^5$. The mass-per-particle of disc and bulge are the same, and similar to that of the halo. The gravitational softening length for both DM and stellar particles is $\epsilon_{\mathrm{DM}} = \epsilon_{\mathrm{S}} = 25 \, \mathrm{pc}$. The opening angle $\theta$ of the tree code has been reduced from $0.7$ used in cosmological simulations to 0.4 for a better quality of force calculation. All models have been run for $10\,\mathrm{Gyr}$, with the angular momentum conservation within 0.5\% and energy conservation within 0.2\%.

\subsection{Initial Conditions}
\label{sec:ICs}

The initial DM component consists of a spherical halo with the modified \citet*[][hereafter NFW]{nfw96} density profile as a function of a spherical radius $r$,
\begin{equation}
    \rho_{\mathrm{h}} (r) = \frac{\rho_{\mathrm{s}} }{[(r + r_\mathrm{c}) / r_\mathrm{s}] (1 + r / r_\mathrm{s})^2} \, \exp\left({- \left(r / r_\mathrm{t}\right)^2}\right),
\end{equation}
where $\rho_{\mathrm{s}}$ is a normalization parameter, $r_{\mathrm{c}} = 1.4 \, \mathrm{kpc}$ is the size of the flat density core, and $r_{\mathrm{s}} = 10 \, \mathrm{kpc}$ is a characteristic radius. We use a Gaussian cut-off radius $r_{\mathrm{t}} = 180$\,kpc to obtain a finite mass.  

The initial conditions of an exponential stellar disk with a density profile are given by
\begin{equation}
    \rho_{\mathrm{d}}(R, z) = \frac{M_\mathrm{d}}{4 \pi R_0^2 z_0} \, \exp\left(-\frac{R}{R_0}\right) \, \mathrm{sech}^2 \left( \frac{z}{z_0} \right),
    \label{eq:rho_d}
\end{equation}
where $R$ is the cylindrical radius, $M_\mathrm{d}$ is the mass of the disk , $R_0 = 2.85 \, \mathrm{kpc}$ is the disk radial scalelength, and $z_0 = 0.6 \, \mathrm{kpc}$ is the disk scaleheight. The disk is truncated at $6 R_{\mathrm{0}} \sim 17 \, \mathrm{kpc}$, i.e., at 98\% of its mass. Velocities of disc particles have been assigned using the epicycle approximation and asymmetric drift correction. The disc dispersion velocities are
\begin{equation}
    \sigma_{R} (R) = \sigma_{R,0} \, \exp \left( -\frac{R}{2R_0} \right)
\end{equation}
\begin{equation}
    \sigma_{z} (R) = \sigma_{z,0} \, \exp \left( -\frac{R}{2R_0} \right),
\end{equation}
where $\sigma_{z,0} = 120\,\mathrm{km/s}$, and $\sigma_{R,0}$ is determined by setting the minimal Toomre parameter $Q$ \citep*[e.g.,][]{Binn08} , to $Q = 1.5$, at $\sim 2R_0$ in all models.

To obtain the initial mass and velocity distributions in spheroidal components, we adopt the iteration method introduced by \citet{rodio06}, see also \citet{rodio09, collier18, li23a}. At each iteration, we freeze the disk and the bulge, and release the DM particles from the initial density distribution for $0.3\,\mathrm{Gyr}$. Next, we assign to each of the unevolved particles, the nearest evolved particle velocity, and iterate.
A total of 70 iterations have been executed until the virial ratio and the velocity distribution of DM particles converge.

For the stellar bulge, we use the Plummer model with a scale length $1\,\mathrm{kpc}$. We utilize the aforementioned iteration method to relax the bulge while the halo and the disc are frozen. About 50 iterations, each lasting $0.03 \, \mathrm{Gyr}$, are implemented until the virial ratio and the velocity distribution of the bulge particles converge.

We use four bulge mass models which differ only by the bulge mass fraction of the total mass, i.e., disk $+$ bulge mass being kept fixed, $B/T$. The bulge models have $B/T = 0$, 0.1, 0.17 and 0.25 bulge mass fractions and are given in Table\,\ref{tab:table1}, where we follow the following notation: the model name, e.g., B10, represents the mass fraction of the bulge, i.e., bulge-to-total mass ratio.  

For the spherical components, the DM halo and the stellar bulge, the angular momenta are zero after the iterations. To spin up the halo and the bulge, we randomly sample the retrograde particles (with respect to the disc rotation axis) in each component, with a constant fraction at all radii, and reverse the azimuthal velocities, $v_{\phi}$, of these particles to achieve the desired spin. In comparison with the bulgeless models in \citet{li23a, li23b}, we obtain spinning halos with $\uplambda = 0.09$ (P90 model) and additional three sets of initially nonspinning  bulge and two initially spinning bulges. For the spinning bulges we have reversed 50\% and 90\%  of originally retrograde particles. 

Two models of DM halos are abbreviated as P00 (for $\uplambda=0$) and P90 (for $\uplambda=0.09$). The bulge models are abbreviated as S50 (nonspinning bulge), S75 (with 50\% of retrograde particles reversed) and S95 (with 90\% of retrograde particles reversed). For example, the model with B10 bulge and its spin S75 immersed in $\uplambda=0.09$ halos is abbreviated as B10-S75-P90.

\begin{table}
	\centering
	\caption{From left to right: DM halo masses, total disk $+$ bulge mass ($T$), bulge mass ($B$), bulge-to-disk mass ratio ($B/D$), and bulge-to-total mass ratio ($B/T$). Each of these four bulge models has been run with three different spins, namely, S50, S75, and S95, as explained in the text. For each of the bulge models, two halo spins have been run, namely, $\uplambda = 0$ (P00 model) and $\uplambda=0.09$ (P90 model).}
	\label{tab:table1}
	\begin{tabular}{lccccc} 
		\hline
		Models  & $M_{\mathrm{h}}$                   & $T$     & $B$  & $ B/D$  & $ B/T$ \\
		          & $\left[ 10^{10}\,\mathrm{M_\odot} \right]$  & $\left[ 10^{10}\,\mathrm{M_\odot} \right]$  & $\left[ 10^{10}\,\mathrm{M_\odot} \right]$\\ 
		\hline
		B0   & 63    & 6.3    & 0        & 0        & 0      \\
		B10  & 63    & 6.3    & 0.63    & 0.11    & 0.10   \\
		B17  & 63    & 6.3    & 1.07    & 0.20    & 0.17   \\
		B25  & 63    & 6.3    & 1.58    & 0.33    & 0.25   \\
		\hline
	\end{tabular}
     \label{tab:tab1}
\end{table}

\begin{figure*}
    \center
    \includegraphics[width=0.8\textwidth]{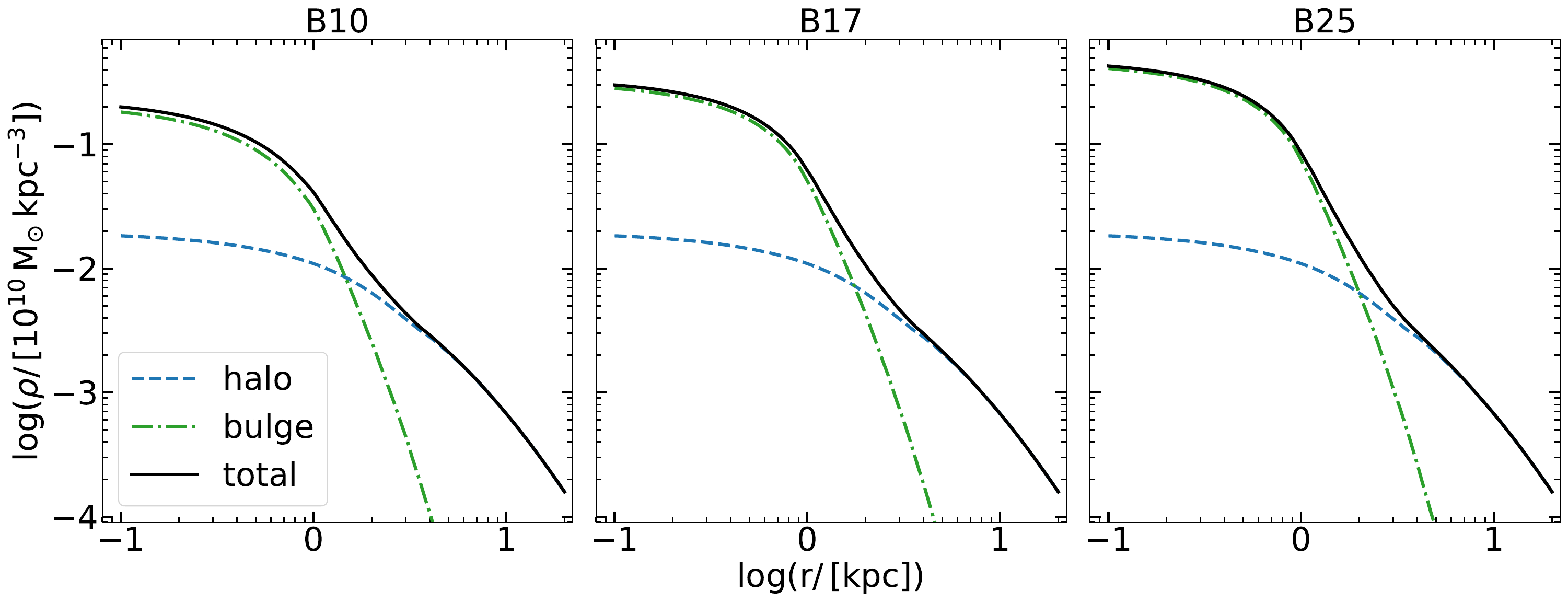}
        \caption{Initial density profiles for the DM halo (the NFW profile, blue dashed line) and the bulge (the Plummer sphere, dashed-dot line) models. also shown is the total mass profile (continuous black line). The disk $+$ bulge masses have been kept the same in all models. }
    \label{fig:rho}
\end{figure*}

\begin{figure*}
    \center
    \includegraphics[width=0.8\textwidth]{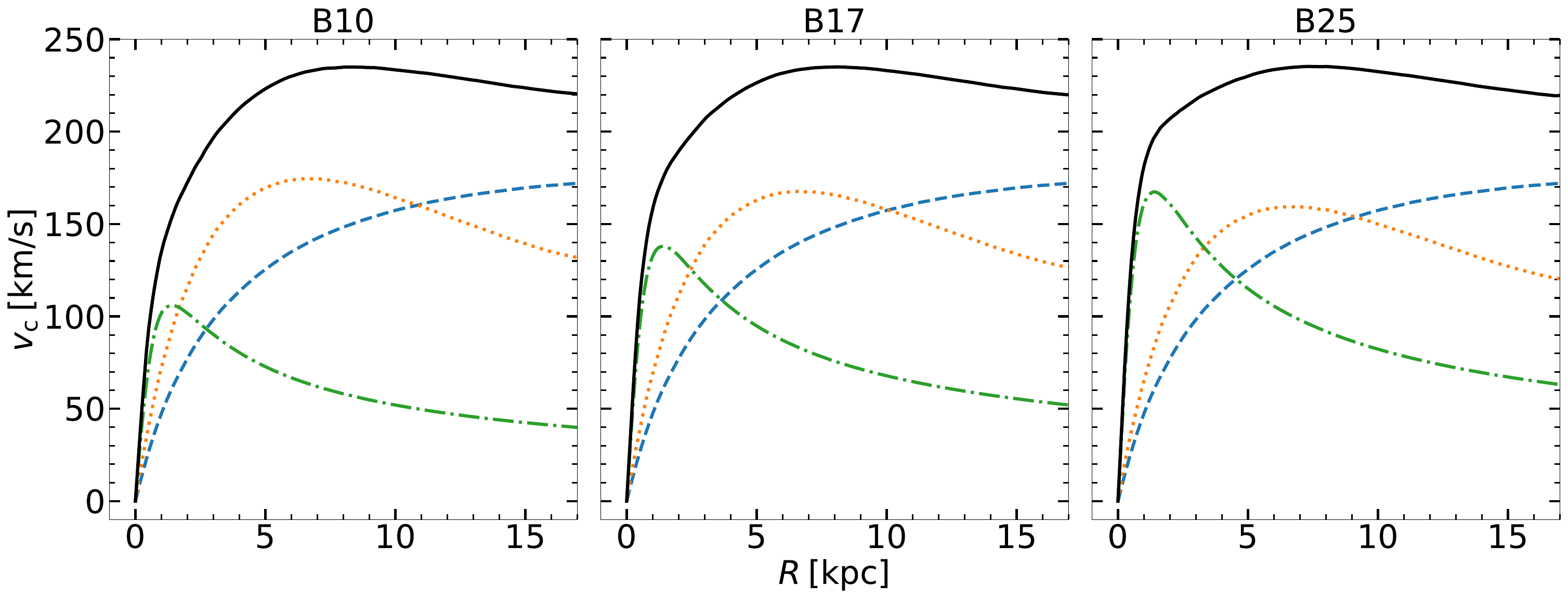}
        \caption{Initial circular velocities for the DM halo  (dashed), stellar disk (dotted), stellar bulge (dashed-dot), and the total (solid) in models with three different bulge masses. From left to right: B10, B17 and B25. Note that the disk $+$ bulge masses have been kept the same in all models.}
    \label{fig:vc}
\end{figure*}

The initial conditions for the density profiles in DM halo, stellar disk and stellar bulges are given in Figure\,\ref{fig:rho}, and the circular velocities in each component are shown in Figure\,\ref{fig:vc}. From left to right the Figures correspond to B10, B17 and B25 bulge models.

\subsection{Maximal angular momentum in bulge-disk-halo systems}
\label{sec:Jmax}

The angular momentum redistribution in the bulge-disk-halo systems is one of the dominant factors in their evolution. Following the work of \citet{sundman1913} (see Appendix), the total angular momentum of any steady collection of particles is bound by the product of the particles moment of inertia and kinetic energy.  
Initially the mass distribution of each of the bulge, disk and halo components are the same for all models. Since the mass distribution fixes the potential energy, the virial theorem fixes the total kinetic energy, but not the components respective 
kinetic energies, so Sundman's inequality may provide different bounds for each models.  We elaborate on this issue in the Appendix, where we use tighter inequalities derived by \citet{pfen19} when the spin vector orientation is known. These  inequalities are tighter than Sundman's because they do no use the $z$ and $v_z$ quantities. 

In Table\,\ref{tab:Jmax}, we calculate the tightest bounds on the maximal angular momentum of the above three components, i.e., the DM halo, disk and bulge, for our models  using Eq.~(\ref{eq:P23}). 
Table\,\ref{tab:Jcomp} presents the initial angular momentum in our models.

\begin{table}
    \centering
    \begin{tabular}{lrrrr} 
    \hline
    Component & B0      & B10     & B17     & B25      \\ 
    \hline
    Halo      & 404.30 & 407.08 & 407.86 & 410.06  \\
    Disk      & 7.38   & 7.05   & 6.76   & 6.36    \\
    Bulge     & 0   & 0.08   & 0.16   & 0.26    \\
    \hline
    \end{tabular}
    \caption{Maximum angular momentum of three components halo, disk, and bulge in three bulge models using Eq.~(\ref{eq:P23}). 
    The units are in $\mathrm{10^{13} \, M_{\odot} \, kpc \, km \, s^{-1}}$.}
    \label{tab:Jmax}
\end{table}

\begin{table}
    \centering
    \begin{tabular}{llrrrr} 
    \hline
    Component              & Spin               & B0        & B10     & B17     & B25      \\ 
    \hline
    \multirow{2}{*}{Halo}  & $\uplambda=0$      & 0.05     & 0.02   & 0.04   & 0.06    \\
                           & $\uplambda=0.09$   & 168.46   & 168.45 & 168.44 & 168.44  \\ 
    \hline
    Disk                   &                    & 6.45     & 6.19   & 5.88   & 5.48    \\ 
    \hline
    \multirow{3}{*}{Bulge} & S50                &           & 0.00     & 0.00   & 0.00    \\
                           & S75                &           & 0.02     & 0.04   & 0.07    \\
                           & S95                &           & 0.04     & 0.08   & 0.13    \\
    \hline
    \end{tabular}
    \caption{Initial angular momentum of halo, disk, and bulge in all models. The units are in $\mathrm{10^{13} \, M_{\odot} \, kpc \, km \, s^{-1}}$.}
    \label{tab:Jcomp}
\end{table}

Since the average velocities do not change much in galaxy components, we see that the maximum angular momentum in the bulge component, which is small both in mass and size, must be small in regard of the disk, and the bulge cannot absorb much angular momentum. For the disk and halo there is a competition between the masses, radii, and velocities, but in view of the larger mass and size of the halo, it is the largest potential angular momentum reservoir by far. 

Plugging in numbers, the bounds on the maximum angular momentum of each component of our models are, in  $\mathrm{10^{13} \, M_{\odot} \, kpc \, km \, s^{-1}}$ units:

\begin{itemize}
\item Bulge: $J < 0.27 $
\item Disk: $J < 7.4 $
\item Halo: $J < 411 $ 
\end{itemize}

The disk is initially near its maximum angular momentum content, so we can expect that most of the angular momentum will migrate from the disk to the halo, because the bulge is unable to absorb much of it.  This does not mean that the bulge is irrelevant.  Since its dynamical time-scale is the shortest of the three components, any angular momentum exchange, even if small in magnitude, can trigger first instabilities like a stellar bar, or a bending, which may propagate  into the larger components, determining the later evolution of the system.   

\subsection{The stellar bar characteristics}
\label{sec:amplitude}

In all models, stellar bars develop from the initially axisymmetric mass distribution. We quantify the bar strength by the Fourier components of the surface density. For $m$ mode, the Fourier amplitude is $\sqrt {a_{\rm m}^2 + b_{\rm m}^2}$, where 
\begin{equation}
    a_{\rm m}(r) = \frac{1}{\pi} \int_{0}^{2\pi} \Sigma(r, \theta) \, \cos(m\theta) \, \mathrm{d} \theta, \; \; \; \; \; m=0, 1, 2, \ldots ,
\end{equation}
\begin{equation}
    b_{\rm m}(r) = \frac{1}{\pi} \int_{0}^{2\pi} \Sigma(r, \theta) \, \sin(m\theta) \, \mathrm{d} \theta, \; \; \; \; \; m=1, 2, \ldots ,
\end{equation}
and $\Sigma(r, \theta)$ is the surface stellar density. To quantify the bar strength, we use the normalized $A_2$ amplitude which is defined as 
\begin{equation}
    \frac{A_2}{A_0}=\frac{ \int_{0}^{R_{\mathrm{max}}} \sqrt {a_{2}^2(r) + b_{2}^2(r)} \, \mathrm{d} r}{ \int_{0}^{R_{\mathrm{max}}} a_0(r) \, dr }.
    \label{eq:barA2}
\end{equation}
We choose the upper limit of integration, $R_{\mathrm{max}}$, as the radius which contains $98\%$ of the disc mass at a given time.

The bar length has been measured using orbital analysis, which is the most reliable method \citep{hell96,bere98,marti06,collier18}. We start by computing the fundamental orbit family, the $\mathrm{x_1}$ family in the notation of \citet{cont80}, which constitutes the backbone of a stellar bar. Next, we determine the extent of this family to the highest value of Jacobi energy. This family is terminated inside the CR. 

\begin{figure*}
    \center
    \includegraphics[width=0.9\textwidth]{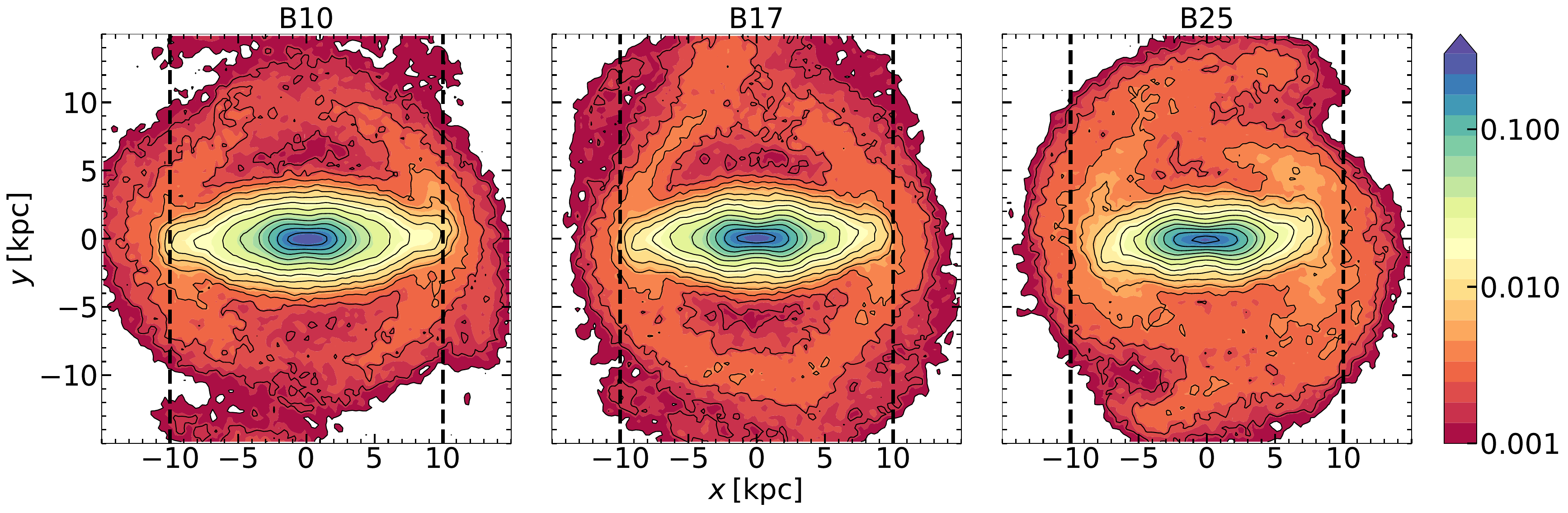}
    \caption{Disk surface density contours for all nonrotating bulge models embedded in the $\uplambda=0$ halo, at $t=6.25\,\mathrm{Gyr}$ when the disk $A_2$ are similar. The color corresponds to the surface density and is given by the logarithmic palette. The isodensity contours are separated by a factor of 1.5 in surface density, and the unit is in $\mathrm{10^{10} \, M_{\odot} \, kpc^{-2}}$. The stellar bar is horizontal and the disk rotates counter-clockwise.The vertical dashed lines have been introduced to compare the bar sizes.}
    \label{fig:disk-bar-in-NRhalo}
\end{figure*}

We measure the vertical buckling strength, $A_{1z}$, i.e. the vertical asymmetry, by calculating the the $m=1$ Fourier amplitude in the $xz$-plane, where the major axis of the bar is aligned with the $x$-axis, and the rotation axis is along $z$-axis, 
\begin{equation}
    \frac{A_{1z}}{A_0}=\frac{ \int_{-x_0}^{x_0} \sqrt {a_{1}^2 + b_{1}^2} \, \mathrm{d} x}{ \int_{-x_0}^{x_0}a_0 \, \mathrm{d} x }.
    \label{eq:barA1z}
\end{equation}
The integral is over the region $|x| < 12$ kpc, $|y| < 3$ kpc, $|z| < 5$ kpc.

The phase of the bar, $\phi_{\rm bar}$, is obtained from
\begin{equation}
\phi_{\rm bar} = \frac{1}{2} \arctan\left(\frac{b_2}{a_2}\right).    
\end{equation}
Generally, $\phi_{\rm bar}$ displays small variations, because of the noise in $b_2$ and $a_2$. We take an average $\phi_{\rm bar}$ in a range of $r$ which defines the bar size.

\subsubsection{Measuring buckling amplitude using an alternative method}
\label{sec:alternative_buck}

The above method of measuring the vertical buckling of stellar bars has provided inconclusive results for some models, due to noise. To remove this ambiguity, we have used a modified vertical asymmetry parameter $A_{\mathrm{asym}}$ introduced by \citet{smirnov18} to measure the buckling strength. The $A_{\mathrm{asym}}$ is defined as
\begin{equation}
    A_{\mathrm{asym}} = \left|\frac{A_2(z>0) - A_2(z<0)}{A_0}\right|
    \label{eq:vasym}
\end{equation}
where $A_2(z>0)$ and $A_2(z<0)$ are the disk Fourier amplitudes, $A_2$, from the {\it numerator} in Equation \ref{eq:barA2} calculated above and below the disk mid-plane, and $A_0$ is calculated from the whole disk. The origin $(x=0, y=0, z=0)$ is set to the center-of-mass of the disk and the $z=0$ plane is perpendicular to the vector of total angular momentum of the disk component within its 90\%-mass-radius.

Our definition of $A_{\mathrm{asym}}$ differs from \citet{smirnov18} which normalized it by $A_2$. This normalization measures $A_{\mathrm{asym}}$ in terms of the bar strength. We find that it is advantageous to measure it in units of $A_0$, which does not vary in the simulations.

\subsection{Measuring the bulge shape}
\label{sec:bshape}

To measure the bulge shape evolution we apply the method of \citet{hell07} by computing the eigenvalues of the moment of inertia tensor for the mass within a specified radius. This determines the axes $a > b > c$ of a uniform spheroid with the same eigenvalues. The ratios of these axes are used to characterize the shape of the bulge, as follows,

\begin{equation}
\frac{b}{a} = \sqrt\frac{e_1-e_2+e_3}{\Delta},
\end{equation}

\begin{equation}
\frac{c}{a} = \sqrt\frac{e_1+e_2-e_3}{\Delta},
\end{equation}
where $\Delta = e_2 - e_1 + e_3$ for the eigenvalues $e_3 > e_2 > e_1$.

\begin{figure*}
    \center
    \includegraphics[width=0.8\textwidth]{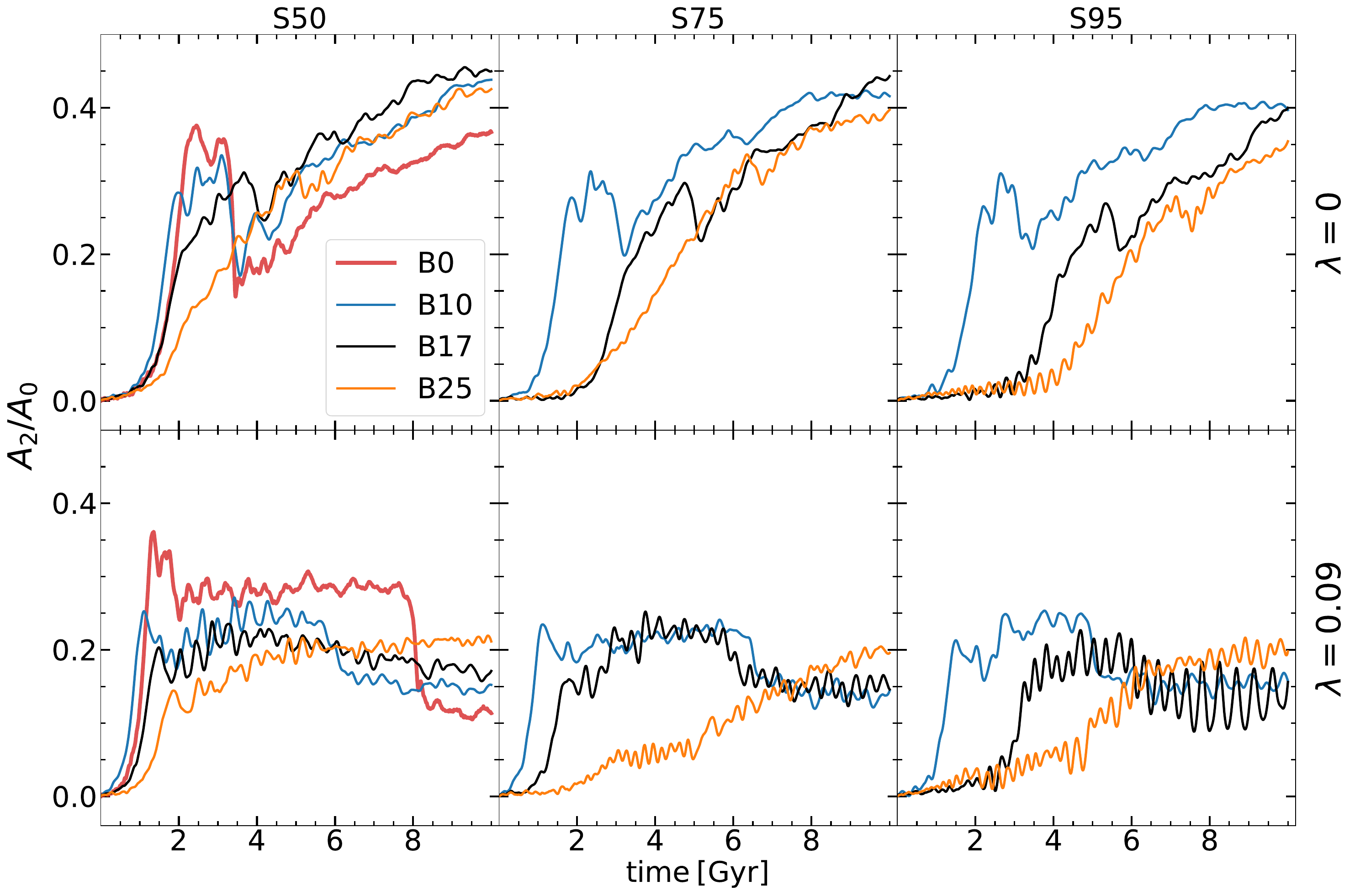}
        \caption{Evolution of the Fourier amplitude $A_{\rm{2}}$ of stellar bars in all models, namely, B0 (red), B10 (blue), B17 (black) and B25 (orange). }
    \label{fig:diskA2}
\end{figure*}

\begin{figure}
    \center
    \includegraphics[width=0.48\textwidth]{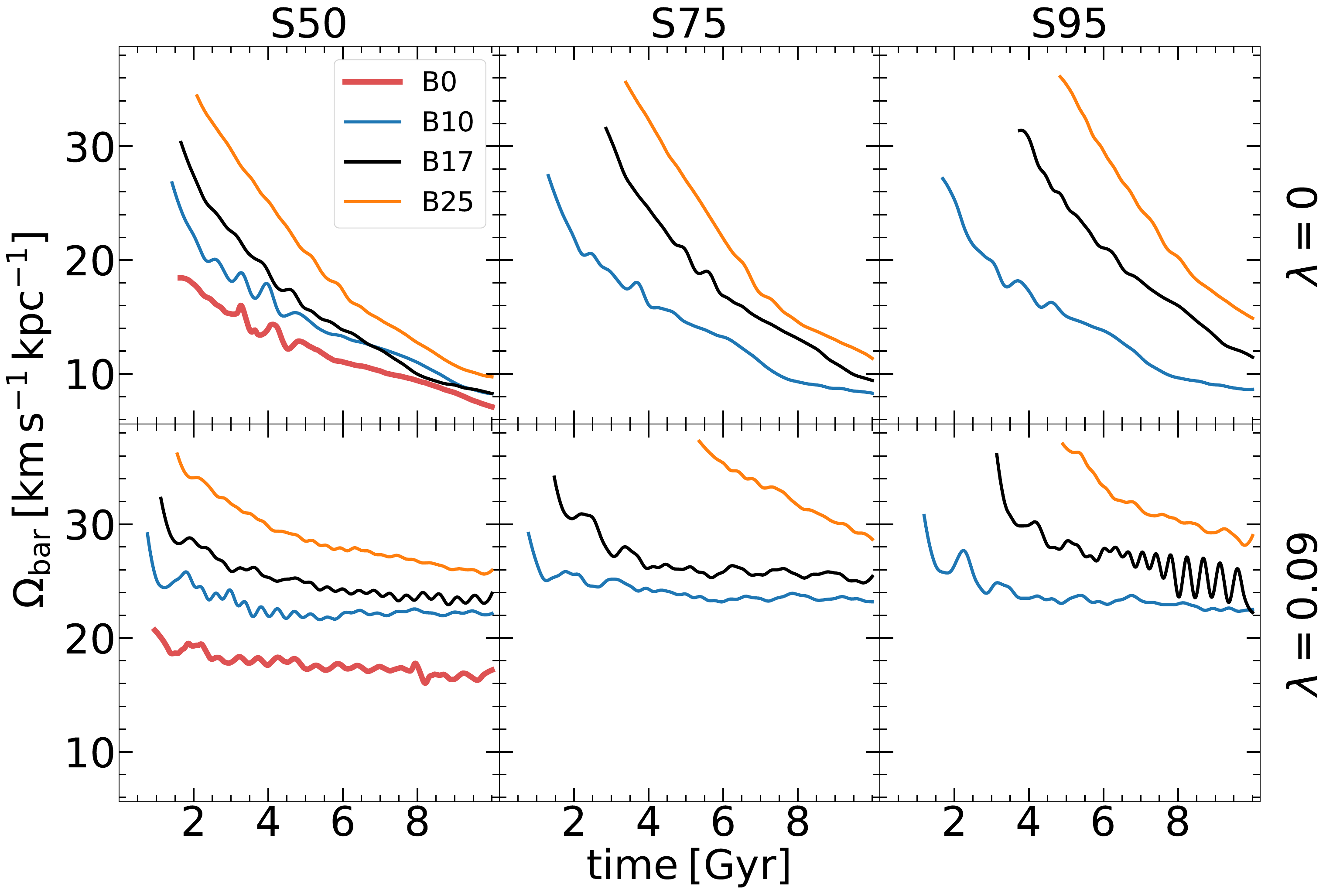}
    \caption{Evolution of the stellar bar pattern speeds, $\Omega_{\mathrm{bar}}$. }
    \label{fig:OmegaBar}
\end{figure}

\begin{figure}
    \center
    \includegraphics[width=0.48\textwidth]{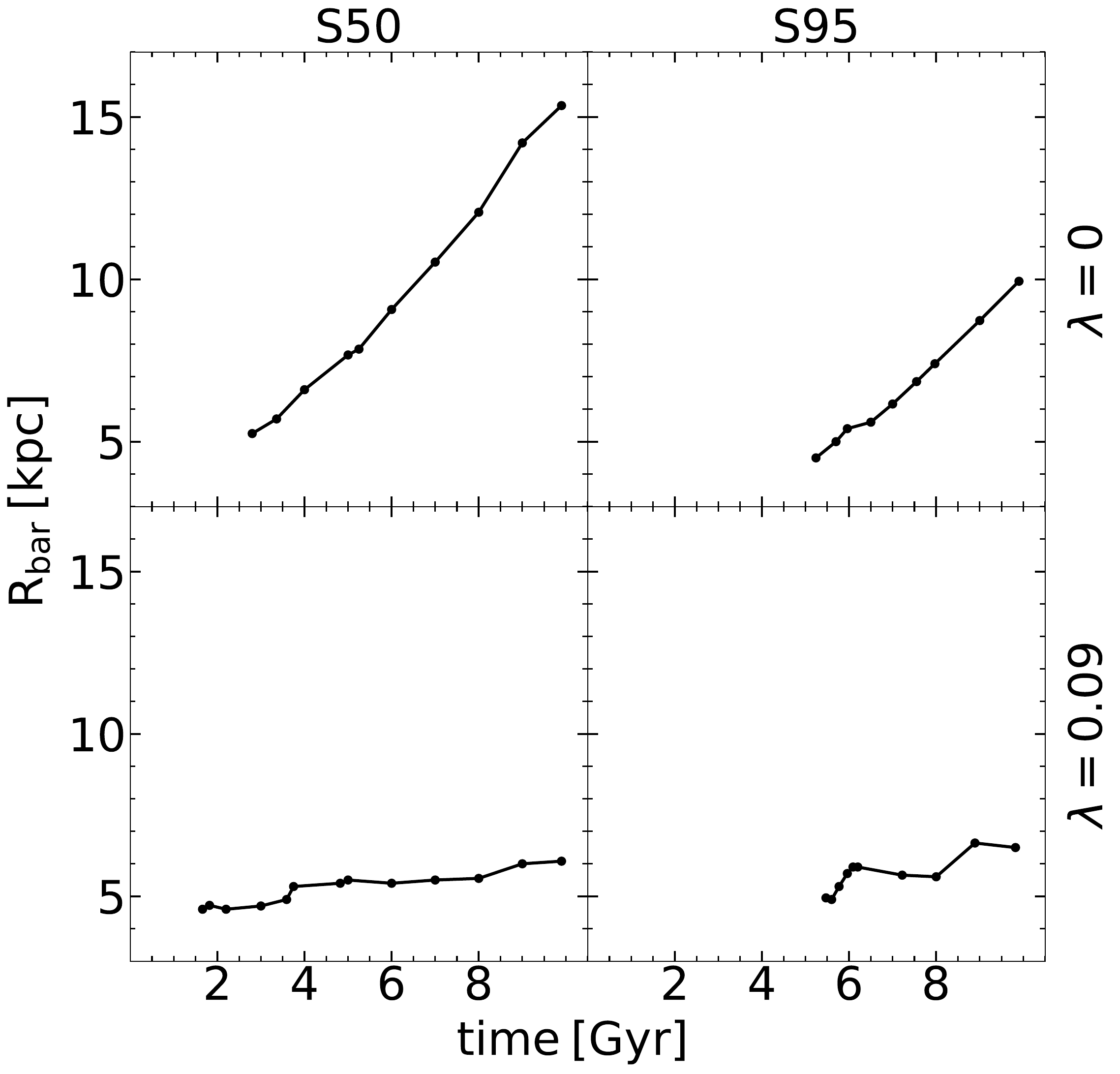}
    \caption{Evolution of bar size of B25 models.}
    \label{fig:barsize}
\end{figure}

\begin{figure*}
    \center
    \includegraphics[width=0.8\textwidth]{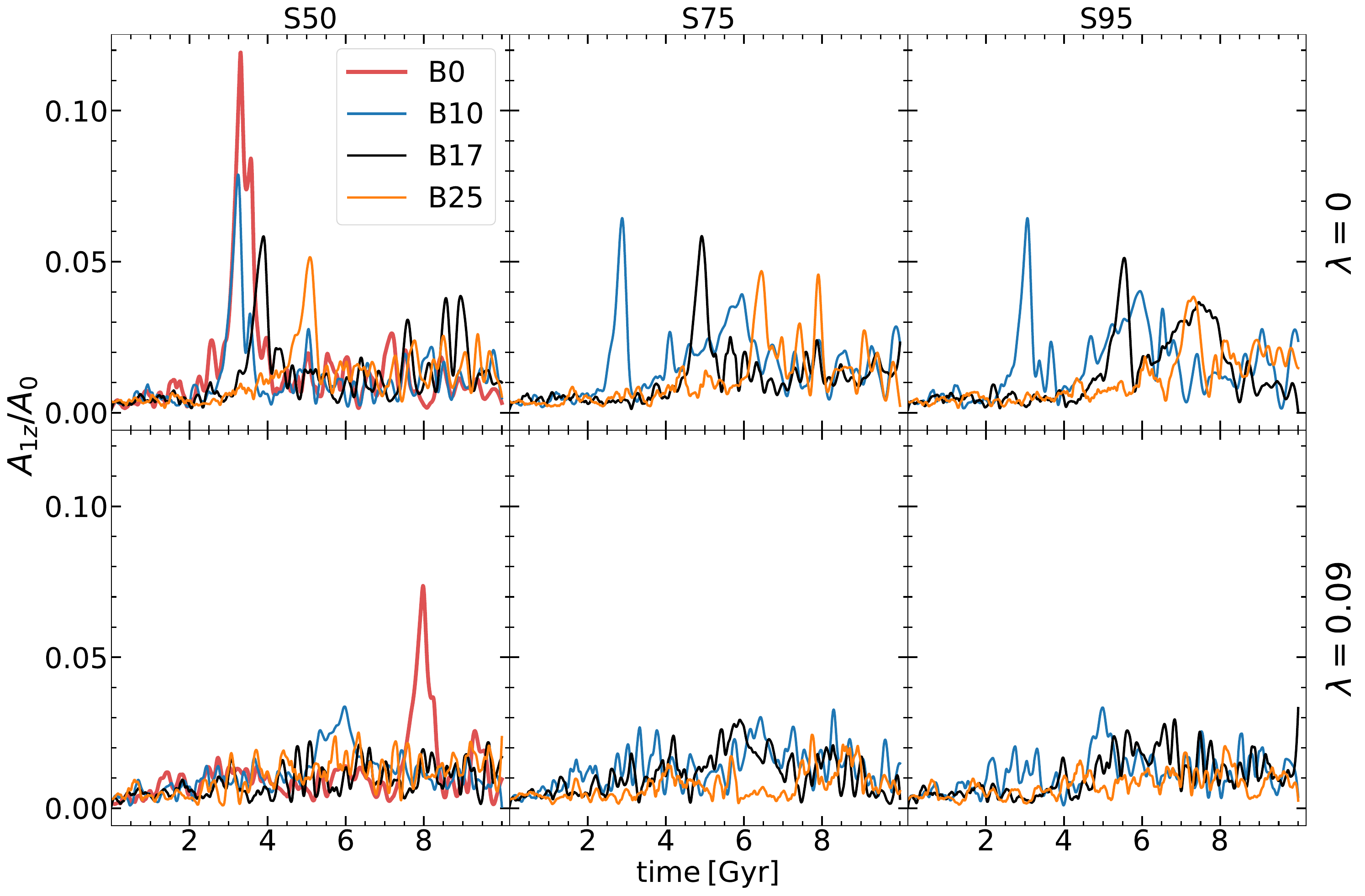}
    \caption{Evolution of the  normalized Fourier amplitude $A_{1z}$ of the disk, using the traditional method.}
    \label{fig:diskA1z}
\end{figure*}

\begin{figure*}
    \center
        \includegraphics[width=0.8\textwidth]{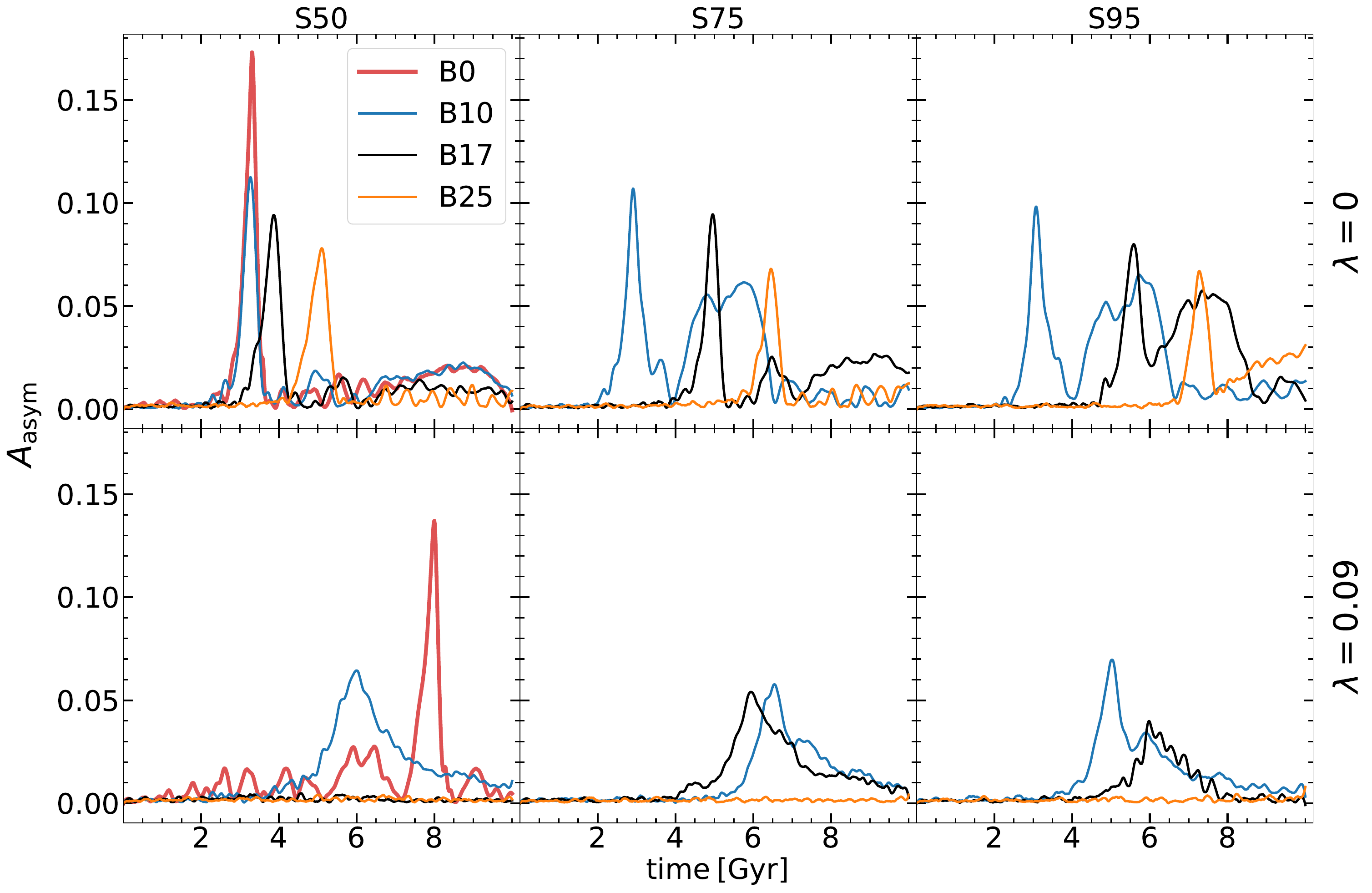}
        \caption{Evolution of vertical asymmetry parameter of the disk in Equation \ref{eq:vasym}, by separating the $z>0$ and $z<0$ surface densities.}
    \label{fig:ver-asym}
\end{figure*}

\section{Results}
\label{sec:results} 

We start by describing the evolution of stellar bars in all the models with the bulge and DM halos with $\uplambda = 0$ and 0.09. As a next step we analyze the bulge evolution for various bulge masses and spin. Finally, we address the interactions between the stellar bars and bulges.

\subsection{Evolution of disk bars in nonspinning halos}
\label{sec:evo-nonspin-halos}

The evolution of stellar disks with nonspinning bulge models immersed in the nonspinning halos is shown in Figure\,\ref{fig:disk-bar-in-NRhalo}, at $t\sim 6.25$\,Gyr, the time when the developed stellar bars have similar strength. The disk bars appear shorter in the more massive bulge models. Also, in all models with bulges in nonspinning DM halos, the surface density in the central few kpc exhibits a dumbbell distribution. The dumbbell does not develop in models with spinning halos of $\uplambda = 0.09$. 

The evolution of stellar bar strength in all our  models is displayed in Figure\,\ref{fig:diskA2}. We start by analyzing the bars in nonrotating halos (top row in this Figure corresponding to $\uplambda=0$ halo). The maximal $A_2$ amplitude before buckling is $\sim 0.3$ and similar in all the models, showing only a slight decrease, less than 10\%, with increasing bulge mass. We also notice that the bar growth is slower for more massive and faster spinning bulges. This is to be compared with $A_2\sim 0.4$ for the bulgeless model. The end product after 10\,Gyr of evolution resulted in strong bars with $A_2\sim 0.35-0.45$.

The growth rate of the stellar bar displays a two-fold effect: it is slower for more massive bulges, and slower for faster spinning bulges as well. The second effect contradicts the growth of stellar bars in DM halos with larger spin, where bars formed earlier with increasing $\uplambda$ \citep[][]{long14,collier18}. 

The bar pattern speeds in nonspinning halos are shown in the top row of Figure\,\ref{fig:OmegaBar} and display a monotonic decay with time. Due to the increased mass concentration from B0 to B25, the initial pattern speeds decrease along this sequence. To avoid the noise, we only calculate the pattern speeds for $A_2 \gtorder 0.1$. Hence, one can observe that the bar instability has been delayed along the B0 $\rightarrow$ B25 sequence as well. 

We have calculated the evolution of bar sizes and present the example in Figure\,\ref{fig:barsize} for B25 models. These sizes are based on the extent of stable $\mathrm{x_1}$ orbits on the characteristic diagrams --- a family of orbits which form the backbones of stellar bars \citep[e.g.,][]{hell96,collier18,li23b}. In this Figure we compare stellar bars in $\uplambda=0$ and $\uplambda=0.09$ DM halos. Note that while bars in the former halos grow monotonically, bars in fast spinning halos essentially grow only till the buckling, and their sizes remain constant after this, in agreement with \citet{collier18}. Of course, bars in fast spinning bulges form later, and hence have a limited time to grow during the runs.

All stellar bar models in {\it nonrotating} halos experience a vertical buckling. This instability is delayed for massive bulges, in tandem with the bar instability itself, and in models with a larger bulge spin (Figure\,\ref{fig:diskA1z}). The buckling amplitude, $A_{1z}$ weakens with increasing bulge mass. 

Some models display the {\it secondary} buckling of stellar bars, which has been first observed in simulations by \citet{marti06}. The secondary buckling is well separated in time from the first buckling. While the first buckling occurs in the central region of a stellar bar, where the harmonic core lies, the secondary buckling happens at larger radii and proceeds much slower than the first one (see Figures\,2, 3 and 5 in Martinez-Valpuesta et al.). The buckling timescales of the first and secondary bucklings are $\sim 0.5$\,Gyr and $2-3$\,Gyr, respectively, so they differ substantially. While secondary buckling is not the subject of this work, we note nevertheless that it can be seen in B10-S75, B10-S95 and B17-S95 bulge models with a nonrotating halo.  

We use the vertical asymmetry parameter, $A_{\rm asym}$, introduced in section\,\ref{sec:alternative_buck} to clarify some prime and secondary bucklings, especially in S75 and S95 bulge models. Figure\,\ref{fig:ver-asym} can be directly compared to Figure\,\ref{fig:diskA1z}. Both types of bucklings can be now clearly verified as the noise has been reduced dramatically in $A_{\rm asym}$. We still observe that the buckling amplitude has decreased with increasing bulge mass. The B0 models exhibit the maximal buckling. What is interesting and was immersed in the noise for $A_{1z}$, is the prolonged timescale of the secondary buckling with $A_{\rm asym}$. In all three models, the length of this buckling is almost 3\,Gyr. In other words, this is the typical extent of the vertical asymmetry in secondary buckling. In fact, this timescale agrees perfectly with \citet{marti06}. 

\subsection{Evolution of disk bar in spinning halos}
\label{sec:evo-spin-halos}

As a next step, we discuss the evolution of stellar bars with bulges immersed in spinning halos.  Figure\,\ref{fig:diskA2} exhibits the evolution of $A_2$ Fourier parameters of the disk. A number of conclusion can be made here. First, the maxima of $A_2$ just before the buckling appear lower than in nonspinning halos, even for the B0 model. This is in accordance with previous simulations \citep{long14,collier18}. With the exception of the B0 model, all models form rather an oval distortion than a bar with $A_2\sim 0.2$, which represents a reasonable boundary between bars and oval distortions \citep[e.g.,][]{bi22b}.

Second, the subsequent evolution of $A_2$ also differs substantially from those in nonspinning halos. The B0 model displays a prolonged plateau after buckling, for about $6.5$\,Gyr, as observed already by \citet{li23b}. Other models which experience buckling show this plateau as well. The length of this plateau decreases with increasing bulge mass. 

Next, the models B0 and B10-S50 are the only ones that experience vertical buckling, at $\sim 8$\,Gyr and 6\,Gyr, respectively. B17 and B25 and the rest of B10 models do not buckle. The B10-S50 buckling amplitude is very low, $A_{1z}\sim 0.035$. The bar in B25 model shows a very slow growth rate and only reaches $A_2\sim 0.2$ late in the run. B25-S75 and B25-S95 reach this value only after 10\,Gyr.

To verify the authenticity of the buckling, we turn to the alternative method described in section\,\ref{sec:alternative_buck}. This method appeared to be more sensitive to the vertical asymmetry as shown by the models in the nonrotating halos 
(section\,\ref{sec:evo-nonspin-halos}). The buckling in B10-S50 has been confirmed (see the lower row in Figure\,\ref{fig:ver-asym}). The $A_{\rm asym}$ parameter also underlines the prolonged width of the buckling in this mode, $\sim 4-5$\,Gyr! 

The alternative method finds a clear buckling asymmetry in B10-S75, B17-S75, B10-S95 and B17-S95, where the standard Fourier method (Eq.\,(\ref{eq:barA1z})) found no buckling at all. Moreover, the width of the buckling is characterized by a long timescale, $\sim 4-5$\,Gyr as well. The amplitude of the buckling is similar in all these models. For the bulge models that experience buckling, this instability develops fast but decays much more slowly than that in the models of nonspinning halos. The B25 models show no buckling in this method as well. Using the alternative method, displays a lower buckling amplitude in spinning halos compared to the nonspinning ones.

We emphasize that the height of the peaks in Figure\,\ref{fig:diskA1z} and Figure\,\ref{fig:ver-asym} have a different meaning.  The $A_{1z}$ Fourier parameter detects the asymmetry in the edge-on disk density distribution. While, $A_{\rm asym}$ parameter detects the difference in the face-on density, between $z > 0$ and $z < 0$. Additional question is why we do not observe the prolonged peaks in $A_{1z}$ while they appear in the $A_{\rm asym}$. This happens because $A_{1z}$ by construction is more sensitive to the central region of the disk, while $A_{\rm asym}$ is equally sensitive at all radii.

In summary, the stellar bar evolution in spinning halos with the mass sequence of bulges differs dramatically from those embedded in the nonspinning halos. The stellar bar amplitude $A_2$ which is substantially lower than for nonspinning halo and lingers around $A_2\sim 0.2$ which represents rather an oval distortion.

\begin{figure*}
    \center
    \includegraphics[width=0.8\textwidth]{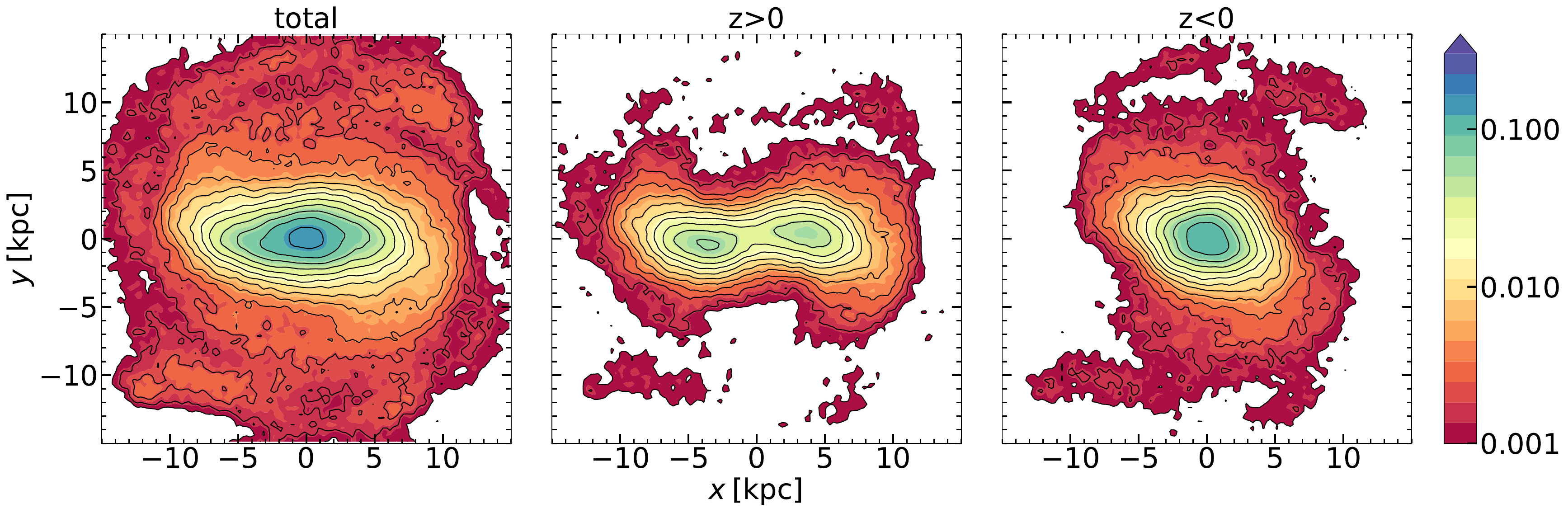}
        \caption{
        {\it Left frame:} The disk surface density contours of the B0 model embedded in the nonrotating halo with $\uplambda = 0$ at the time of maximal buckling, $t = 3.35$\,Gyr.  {\it Middle and right frames:} same as left frame but separating $z > 0$ and $z < 0$ surface densities. The $z=0$ plane is defined in Eq.\,(\ref{eq:vasym}) and the color palette is in the same level as Figure\,\ref{fig:disk-bar-in-NRhalo}.}
    \label{fig:faceon-asym-b0}
\end{figure*}

 \begin{figure*}
    \center
    \includegraphics[width=0.8\textwidth]{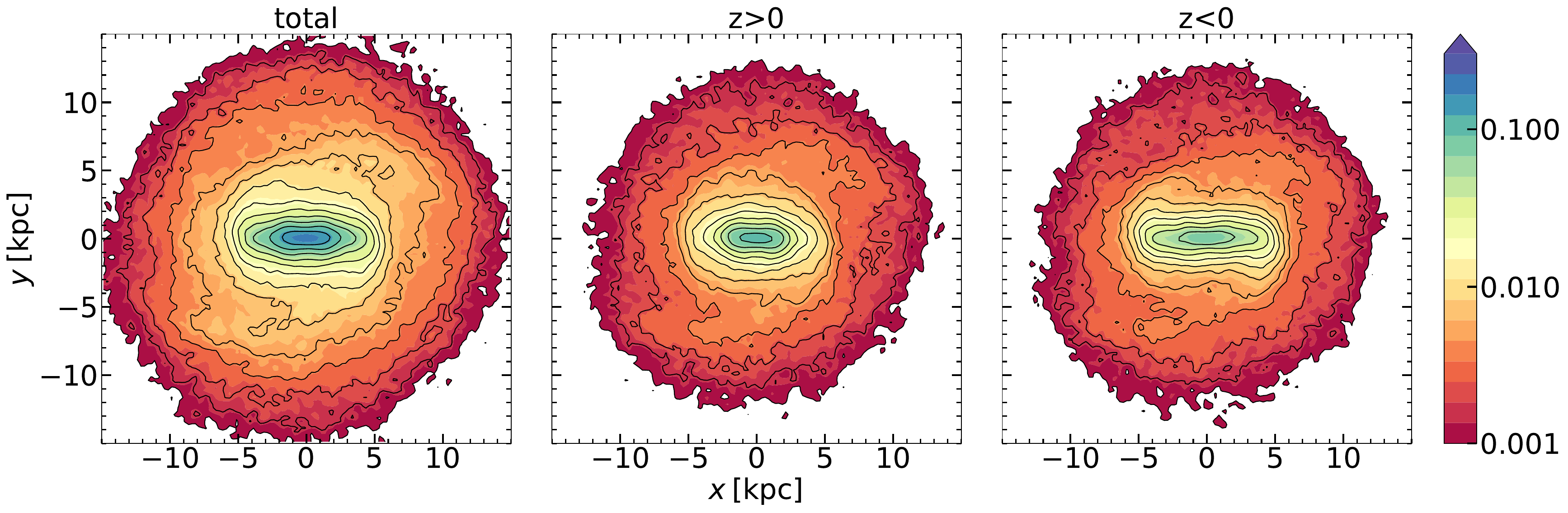}
        \caption{Same as in Figure\,\ref{fig:faceon-asym-b0}. But for the B17-S75 model embedded in the spinning halo with $\uplambda=0.09$ at the time of maximal buckling, $t = 6.00$\,Gyr.}
    \label{fig:faceon-asym-b17}
\end{figure*}

\begin{figure}
    \centering
    \includegraphics[width=0.48\textwidth]{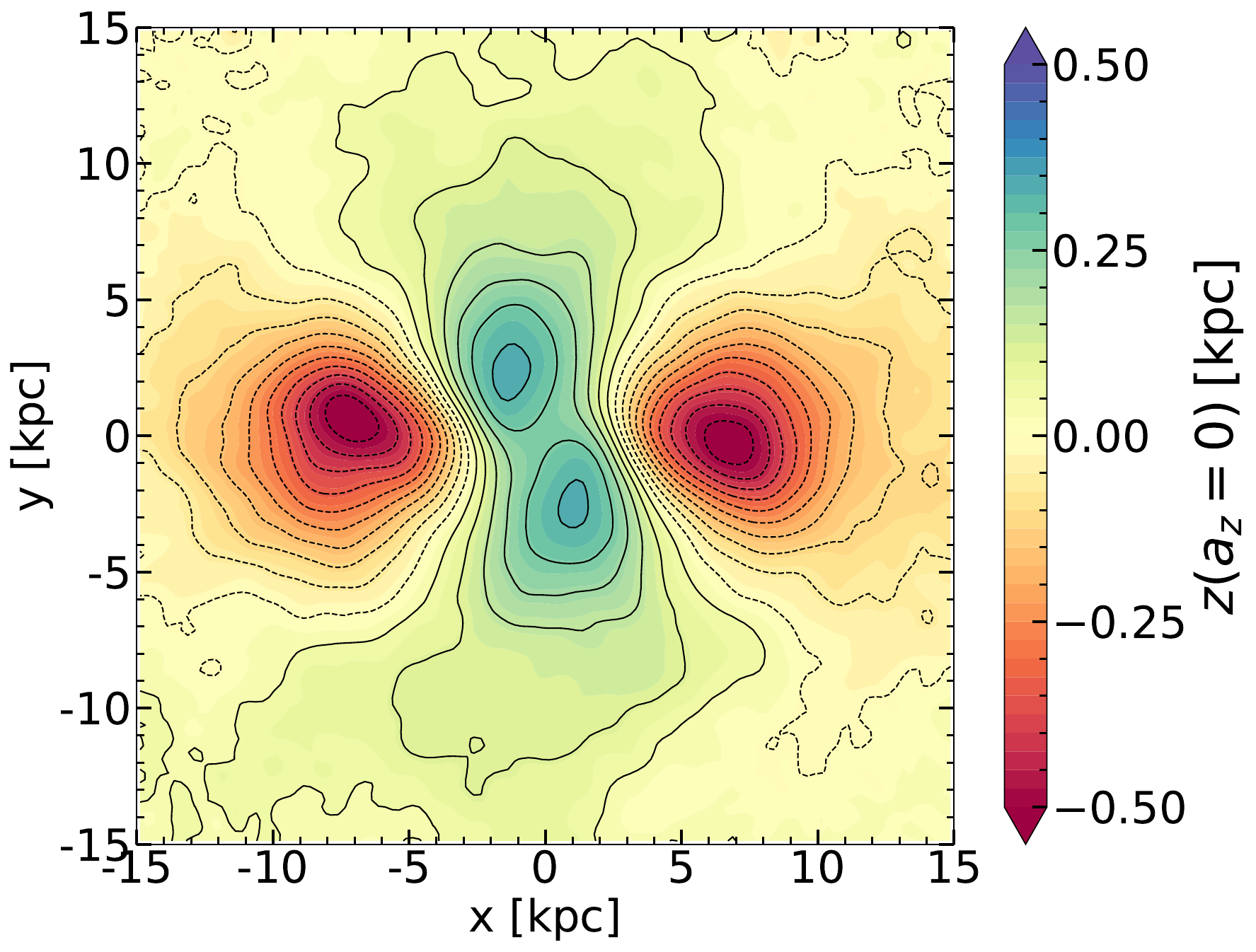}
    \caption{The face-on Laplace plane for the model B10-S50 with the $\uplambda=0$ halo, at the time of the maximum buckling,  $t=3.30 \: \mathrm{Gyr}$. The major axis stellar bar is horizontal. During this buckling, the stellar bar center bends upward.}
    \label{fig:az0_B10P00_1}
\end{figure}

\subsubsection{Vertical asymmetry of stellar bars using the alternative method}
\label{sec:smirnova}

\begin{figure*}
    \center
        \includegraphics[width=0.8\textwidth]{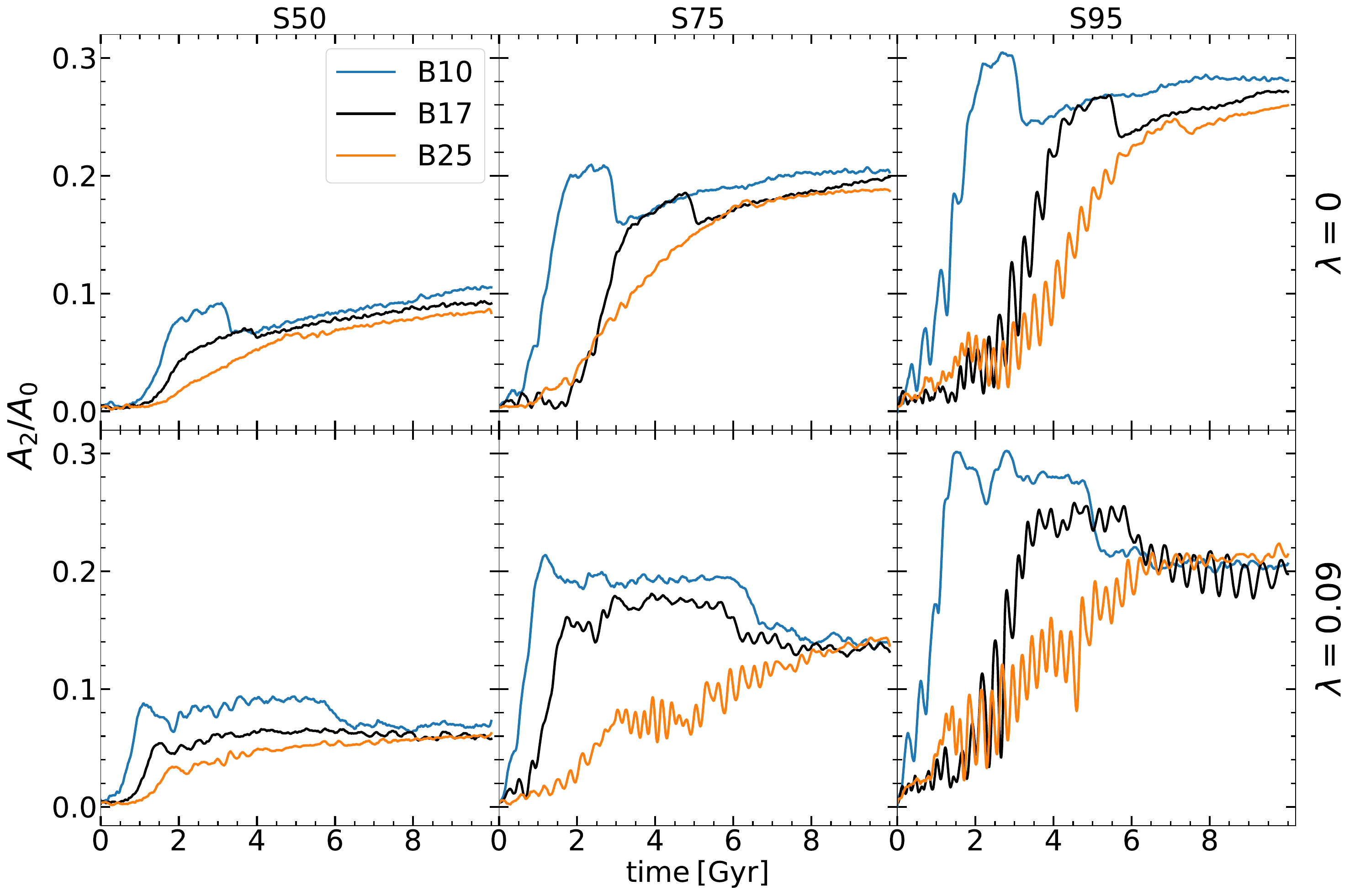}
        \caption{Evolution of the Fourier amplitude $A_{2}$ of the bulge component normalized by the monopole term $A_0$. The top row: disk $+$ bulge immersed into nonrotating halo, with $\uplambda = 0$; Bottom row: immersed into spinning halo with $\uplambda = 0.09$. Left column: nonspinning bulges (S50 models); Middle column: intermediate spin bulges (S75 models); Right column: Fast spinning bulges (S95 models).}
    \label{fig:bulgeA2}
\end{figure*}

\begin{figure*}
    \center
        \includegraphics[width=0.8\textwidth]{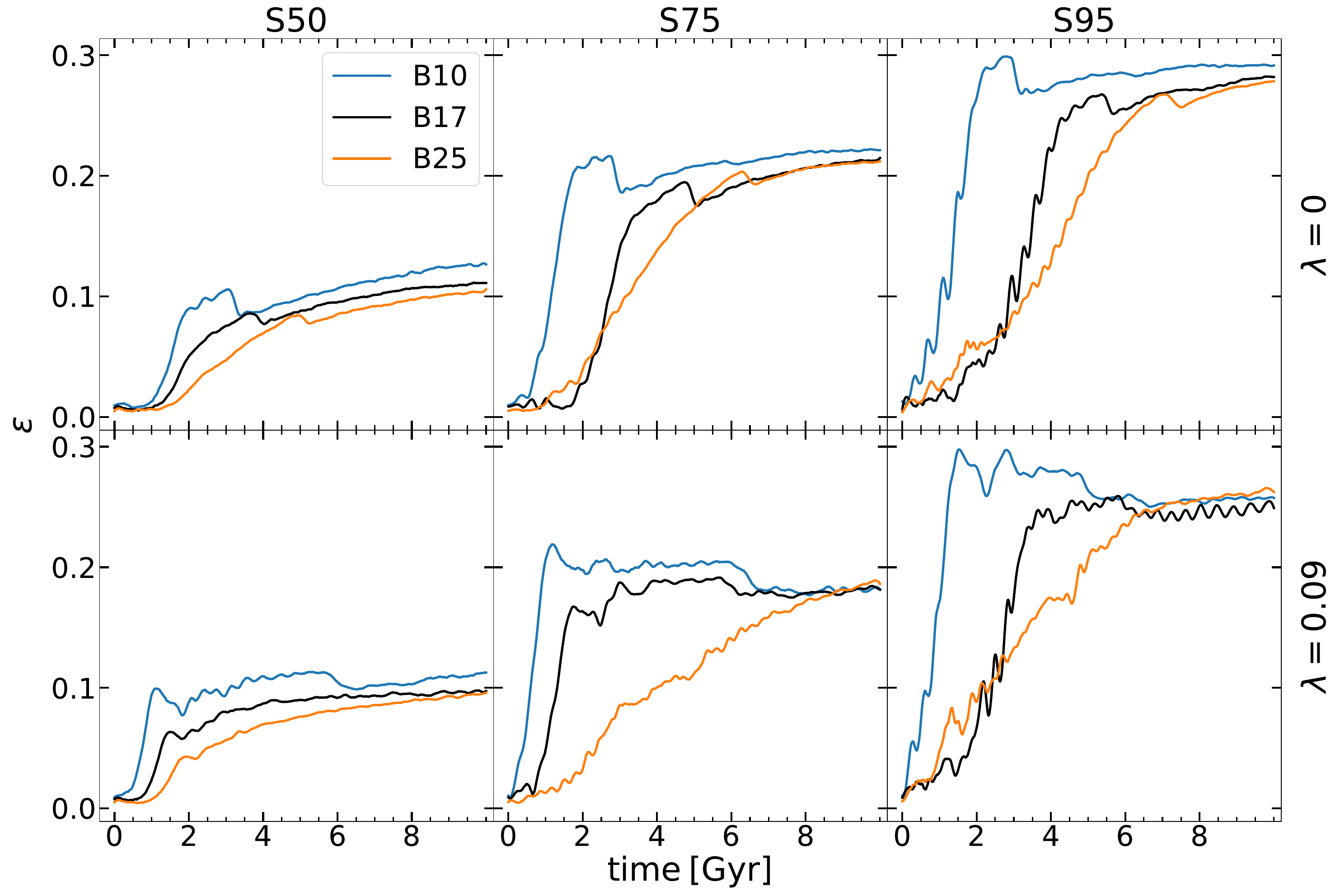}
        \caption{Evolution of ellipticity, $\epsilon = 1-c/a$, of the bulge edge-on isodensity contours within $R<5 \: \mathrm{kpc}$.}
    \label{fig:eccentr}
\end{figure*}

To understand and visualize the alternative method to determine the vertical asymmetry more fully, we present the face-on snapshots of the disk for all the models with and without the bulge. Figure\,\ref{fig:disk-bar-in-NRhalo} displays the face-on stellar disks and bulges for B10-S50 to B25-S50 models immersed in nonrotating DM halos. All the snapshots are given at $t\sim 6.25$\,Gyr, after the stellar bars have reached the amplitude close to the maximal one.  

We test the method of \citet{smirnov18} for sensitivity to the vertical buckling. The traditional method of measuring the buckling amplitude $A_{1z}$ has difficulty to detect the instability in some of the models with nonspinning and spinning halos. Using this alternative method, the noise is dramatically suppressed. This allows to recognize the buckling process more clearly. For example, for the bulgeless model B0 with a spinning halo, $A_{1z}\sim 0.075$, while the alternative method results in $\sim 1.25$ in Figure\,\ref{fig:ver-asym}. Moreover, the B10 nonspinning bulge within the halo of $\uplambda=0.09$, has $A_{1z}\sim 0.035$, while the new method results in 0.65 buckling amplitude. Factor of 20 higher leads to a clear detection of the buckling in B10 and B17 intermediate and high rotating bulge. Not only the buckling amplitude is increased in the new method, but also the duration of the buckling points a slowly restoring vertical symmetry over a few Gyr due to a higher sensitivity of the method.

Most interesting is observing the actual response of the disk to the buckling in Figure\,\ref{fig:faceon-asym-b0}, for a bulgeless B0 model in a nonrotating halo. While the left frame shows the usual surface density distribution with a fully developed stellar bar, the middle and right frames display the separated disks for $z > 0$ and $z < 0$ surface densities. The $z > 0$ frame differs from the full surface density in the left frame by exhibiting a dumbbell shape in addition to having a very elongated distribution. At the same time, the $z < 0$ frame displays no dumbbell and being inclined by about $30^\circ$ to the bar which is horizontal. Its ellipticity is much smaller than that of the dumbbell frame. In this Figure, the bar is horizontal and the disk rotates counter-clockwise. 

One can understand the appearance of the dumbbell shape during downward buckling --- such a buckling is always associated with a double maxima response in the opposite direction, i.e., formation of the bell-shape and a dumbbell. As an example, we point to the image of an edge-on disk shown at the time of buckling downwards at $t = 2.4$\,Myr in Figure\,3 of \citet{marti06}. What is new and unexpected is that the response above and below the midplane of the disk different orientation in the $x-y$ plane. Based on the bar position and the disk rotation, we conclude that the response of the $z < 0$ part of the disk is trailing the bar and the response of the upper hemisphere.

Such a response will induce the shear across the disk  midplane. This phenomenon has been indeed observed by \citet{li23a} in their Figures\,6--9. They have observed formation of circulation cells and the associated vorticity in response to the shear. The associated stellar orbits giving rise to this circulation have presented in the Appendix of Li et al. work, and also discussed by \citet{hell96}. 

Figure\,\ref{fig:faceon-asym-b17} describes the same phenomenon for the B17-S75 model embedded in the $\uplambda = 0.09$ halo. Here we do not observe the misaligned response between the upper and lower disk hemispheres. The difference with the Figure\,\ref{fig:faceon-asym-b0} is that now a B17-S75 bulge is present and the DM halo has a spin. The amplitude of the asymmetric response is much lower, $A_{\rm asym}\sim 0.6$ compared to $\sim 1.35$ in the B0 model. Hence, the bulge acts to damps this asymmetry and this damping ability is expected to be stronger for more massive bulges. 

We have tested the appearance of the shear and the associated phenomena for all models, but show only two models described above. To summarize these results, the shear and misaligned response has been detected in all models with nonspinning DM halos. Only B10-S50 models with $\uplambda = 0.09$ halo exhibits this behavior. The rest of the models do not display it. The presence or absence of this behavior corresponds to appearance of $A_{\rm asym}$ peak in Figure\,\ref{fig:ver-asym}. The $A_{1z}$ Fourier coefficient which is typically used to measure the buckling strength of stellar bars is not so sensitive to buckling.

Using the $A_{\rm asym}$ amplitude can be supplemented by mapping of the Laplace plane for the barred disk \citep[e.g.,][]{dekel83,Binn08}. In Figure\,\ref{fig:az0_B10P00_1}, we present an example of B10-S50 model during the maximal buckling, when the twin negative peak formed along the stellar bar (horizontal), and a positive peak along the minor axis. This configuration is supported by the formation of a circulation cell of compression/stretching during the buckling analyzed in \citet{li23a}.

We do observe that the $z > 0$ forming a bar is shorter than the $z < 0$ response, about 5\,kpc versus 6\,kpc radius bar, respectively. 

\begin{figure*}
    \center
        \includegraphics[width=0.9\textwidth]{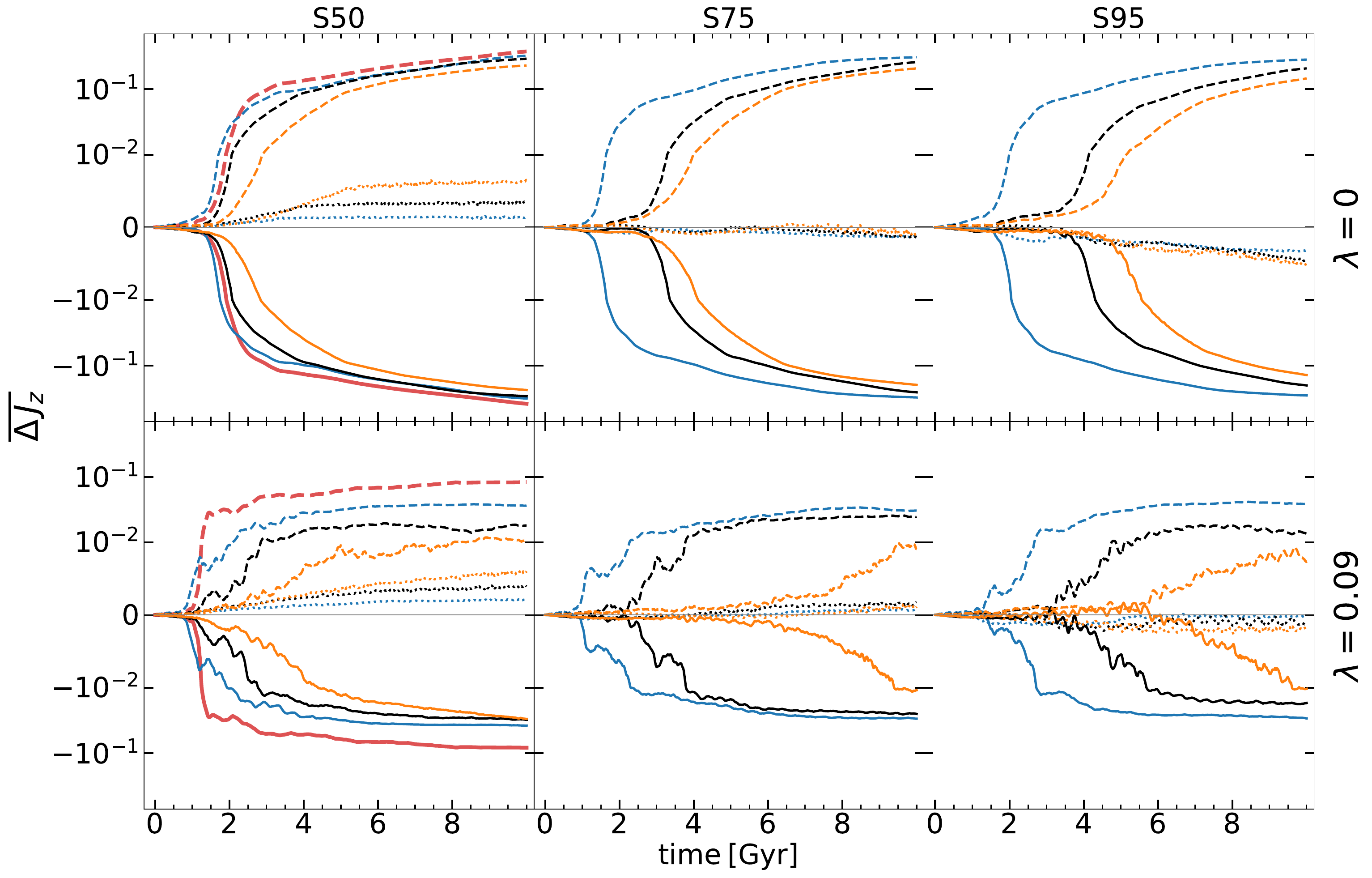}
        \caption{Evolution of the angular momenta exchange, $\overline{\Delta J_{z}}$ in Equation\,\ref{eq:rel_Jz}, of each component, namely, the DM halo (dashed lines), disk (solid lines), and bulge (dotted lines), normalized by the disk angular momentum of the  bulgeless model B0 at $t=0$\,Gyr, in B0 (red), B10 (blue), B17 (black), and B25 (orange) models. The upper row displays the models with $\uplambda = 0$ halos and the bottom row shows models immersed in $\uplambda = 0.09$ halos. \textit{Note that the $y$-axis is in linear scale in the range $[-10^{-2}, 10^{-2}]$, otherwise in logarithmic scale.}}
    \label{fig:Jz-all}
\end{figure*}

\subsection{Evolution of the bulge in the presence of stellar bar}
\label{sec:bulge_resp}

We turn to the bulge models, both nonspinning and spinning initially. We start by discussing the evolution of their $m=2$ modes in $xy$-plane. For the nonspinning bulges, this mode grows in response to the stellar bar development. But for bulges with an initial spin, the $m=2$ mode can be triggered spontaneously, at least in principle, and evolve not in isolation but in interaction with the large-scale stellar bar. In all cases, the bulge and the bar possess gravitational quadrupole moments, and therefore can exchange their angular momenta both between them and with the host DM halo, again in principle.

Figure\,\ref{fig:bulgeA2} displays the evolution of the bulge $A_2$ Fourier coefficient for bulges embedded in $\uplambda = 0$ nonspinning halos (top row) and $\uplambda = 0.09$ spinning halos. For nonspinning bulges, the amplitude of $m=2$ remains low, $A_2 \ltorder 0.1$, which in our notation would correspond to the oval distortion. 

The bulge $A_2$ grows in response and completely mimics the development of the bar instability in the stellar disk (left column of Figure\,\ref{fig:bulgeA2}, S50 models). It responds by weakening during the buckling of the parent stellar bar (if this occurs). In nonrotating halo, the maximal bulge $A_2$ barely reaches $\sim 0.1$, while in the spinning halo, it is even weaker, again mimicking the evolution of the parent stellar bar.   

For the middle column, the bulge S75 models, develop a much stronger $m=2$ mode. Yet, it can be barely considered a "bulge bar" because its amplitude is $\sim 0.2$. Again it seems that the bulge $m=2$ mode has been induced by the parent bar instability in the stellar disk. Finally, the right column displays development of a strong bulge bar in S95 bulge models, reaching the amplitude of $\sim 0.3$.

In the $xy$-plane the bulge responds to the stellar bar, but in the vertical plane, it flattens in response to rotation and the disk potential (Figure\,\ref{fig:eccentr}).  Thus the resulting Figure of a spinning bulge is that of a triaxial ellipsoid. This flattening is usually measured by ellipticity, i.e., $\epsilon = 1 - c/a$, where $c$ is directed along the $z$-axis, and $a$ and $b$ are the semimajor and semiminor axes in the $xy$-plane (see section\,\ref{sec:bshape} for our method to calculate these axes using the moment of inertia of the bulge).

\subsection{Angular momentum redistribution in the disk-bulge-halo system}
\label{sec:evo-J}

\begin{figure*}
    \center
        \includegraphics[width=0.98\textwidth]{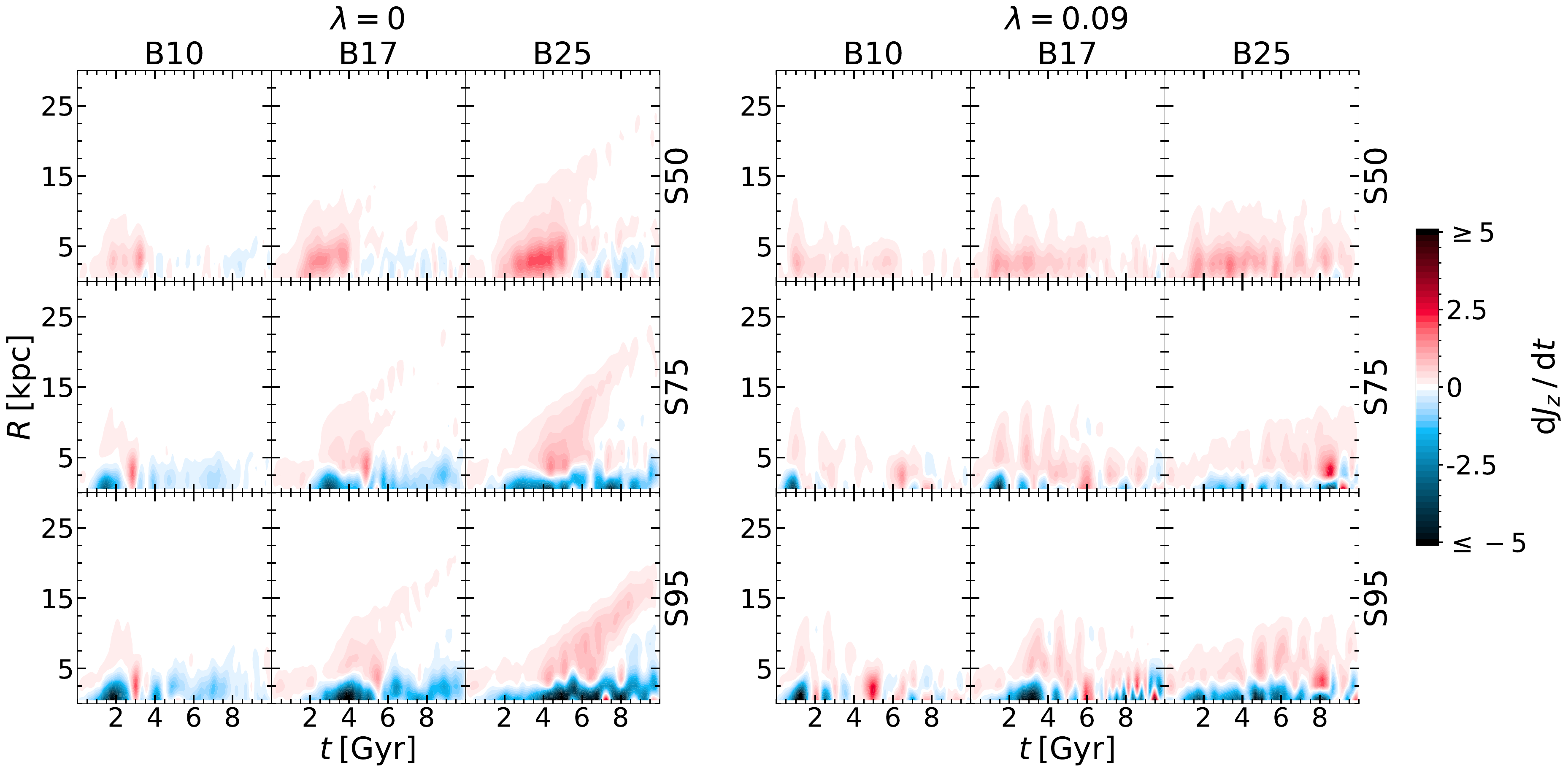}
        \caption{The rate of the angular momentum change in the bulge, $\dot J_{z}$, in the models with $\uplambda = 0$ halo (left) and $\uplambda = 0.09$ halo (right). $\dot J_z$ is calculated by every 0.05\,Gyr within cylindrical shells $\Delta R=1\,\mathrm{kpc}$ and $|z| < 6\,\mathrm{kpc}$. The red and blue color palette corresponds to the gain and loss of angular momentum, respectively. $\dot J_{z}$ is given in units of $\mathrm{10^{10}M_{\odot}\,kpc^2\,Gyr^{-2}}$.}
    \label{fig:Jdot}
\end{figure*}

As a next step we analyze the redistribution of angular momentum between the three components, namely, the bulge, disk and parent DM halo. We start by following the evolution of angular momentum in each of these components, then following the resonant interactions between them.

Figure\,\ref{fig:Jz-all} exhibits the angular momentum gain and loss by each of the three morphological components, namely,

\begin{equation}
    \overline{\Delta J} = \frac{J(t) - J(t=0)}{J_{\rm B0,disk}(t=0)} 
    \label{eq:rel_Jz}
\end{equation}

We have normalized the angular momenta of each component by the initial angular momentum of stellar disk, $J_{\rm B0, disk}(t=0)$, of the bulgeless model B0. 

In all cases, the lion share of angular momentum redistribution happened between the disk and its parent halo. The former lost and the latter gained the angular momenta, i.e., the stellar disk is the source of the angular momentum in the system and the halo is its sink. This exchange is much larger in the models with $\uplambda = 0$ halos (top row in this Figure), by a factor of $\sim 10$ compared to models with spinning halos, $\uplambda = 0.09$. As formation of stellar bars requires a substantial loss of angular momentum in the bar region, its transfers mostly to the parent halo. 

The decreased angular momentum transfer between the disk and the halo explains in part the weaker bars forming inside such halos (e.g., Figure\,\ref{fig:diskA2}). In addition, this Figure also points to a difference between the bar strength in the bulgeless B0 model and B10-B25 models, and therefore to an additional angular momentum transfer, this time between the stellar disks and their bulges.

Our next question is to determine the role of the bulge in the angular momentum redistribution. All nonspinning bulges in $\uplambda = 0$ halos gain angular momentum, but this amount is small and on the linear scale compared to the angular momentum gained by the halos on the logarithmic scale. All the bulges inside $\uplambda = 0$ halos gain angular momentum until the associated stellar bars have reached the maximal amplitude at the buckling. After this, the transfer of angular momentum has almost ceased until the end of the run at 10\,Gyr. For initially nonspinning bulges immersed inside $\uplambda = 0.09$ halos, the bulges continue to gain angular momentum till the end of the simulations, and the saturation effect is not visible almost at all.

Overall, all initially nonspinning bulges gained angular momentum by a factor of $\sim 50$ less than their parent halos. Interesting that more massive bulges in this left column of Figure\,\ref{fig:Jz-all} have gained more angular momentum.
We have checked and found that this result holds for the specific angular momentum of bulge particles as well.

For initially intermediate spin bulge, i.e., S75 models, the change of angular momentum is basically non-existent, independent of the bulge mass. For the initially fast spinning bulges, i.e., S95 models, in $\uplambda = 0$ halos, the bulges appear to lose their angular momentum, although compared to the parent stellar disks, this amount is not significant. The same bulges in $\uplambda = 0.09$ halos, in contrast, lose very little angular momentum.

Based on our understanding of the source and sink of angular momentum in the bulge-disk-halo system, we turn to the rate of the angular momentum transfer to and away from the bulge, attempting to answer the question, what is the role of the bulge in this evolution. We use the method developed by \citet{villa09}, see also \citet{collier18} and \citet{li23b}. The cylindrical radius has been binned into cylindrical shells with the height of $|z| = 6$\,kpc. We calculate the angular momentum change in these shells, $J(t)$, and obtain the rate of change, $\dot J = [J(t+\Delta t)-J(t)]/\Delta t$. 

The left panel in Figure\,\ref{fig:Jdot} exhibits the rate of change of $J$ in the all the bulge models immersed in the $\uplambda = 0$ halos. It must be compared with Figures\,\ref{fig:bulgeA2} and \ref{fig:Jz-all}. The B0 model can be found in \citet{li23b}. We observe that initially nonrotating bulges (S50 models) absorb the angular momentum almost all the time. After the associated stellar bar buckling, the angular momentum absorption rate decreases sharply and some emission of $J$ appears. This explains that the termination of the bulge spinup saturates in Figure\,\ref{fig:Jz-all}. 

For S75 models, the $J$ redistribution differs. The inner bulge emits its angular momentum at small radii and absorbs it at larger radii --- as a result almost no net $J$-transfer occurs. This absorption increases even more in S95 models, leading to the loss of $J$ by these bulges.

The right panel in Figure\,\ref{fig:Jdot} displays the same phenomenon in $\uplambda = 0.09$ halos. Pure spin-up of bulges in S50 models, appearance of absorption in the inner bulges in S75 models, and dominant absorption in S95 models.

\subsection{Orbital spectral analysis}
\label{sec:osa} 

Finally, we performed the orbital spectral analysis of the bulge-disk-halo systems, with a particular emphasis on the bulge, in order to examine the role of resonances in angular momentum transfer among the morphological sub-components.  We followed the method developed and used by us \citep{marti06,dubi09,collier19a}. The dimensionless frequency ratio is  defined as $\nu = (\Omega - \Omega_{\rm bar})/\kappa$, where $\Omega$ and $\kappa$ are the angular and epicyclic frequencies characterizing the individual orbits. The frequency ratio $\nu$ is binned in $\Delta\nu = 0.005$, and we choose a random sample of 50\% of bulge particles from each model for this analysis. In this notation, the main resonances are the inner Lindblad resonance (ILR), the corotation resonance (CR), and the outer Lindblad resonance (OLR). They correspond to $\nu = 0.5$, $0$, and $-0.5$, respectively.

Our goal is to find the fraction of bulge orbits trapped at particular resonances and whether these resonances gain or loose angular momentum. The orbital spectral analysis has been performed at different times before and after buckling of stellar bars in models with a nonrotating bulge and nonrotating halo. This allowed the measure of the angular momentum gain or loss at these resonances. The resulting distribution of bulge particles with $\nu$ has shown that the main resonances which trap these particles are $\nu = 0.5$, $\nu = 0$, and $\nu=-0.5$.

We have calculated the trapping efficiency and angular momentum gain/loss for initially nonrotating bulges of B10, B17 and B25 embedded in $\uplambda=0$ halos. In these models, the bulge ILR, has always shown absorption of angular momentum, with an efficiency of trapping increasing with the bulge mass fraction from $\sim 14\%$ to $\sim 24\%$. This is opposite to what we observe for the barred disks which always emit angular momentum at this resonance. The second resonance corresponds to the OLR with a trapping efficiency $\sim 10\%$ to $\sim 15\%$, but it involves the retrograde orbits with $\Omega < 0$. These orbits have shown an absorption of negative angular momentum. To understand this behavior, we looked at these orbits and found that they lie deep within the bulge and circulate in the direction opposite to that of the stellar bar and disk. They appear elongated orthogonal to the bar. 

We have also detected small trapping of bulge orbits at $\nu = -1$, which corresponds to the Lagrange points. But the efficiency of trapping by this resonance has been low, $\sim 1\%$, and we did not follow it. In comparison, \citet{saha12} recorded an absorption of angular momentum at this resonance, although with a varying efficiency.

 In our analysis, the CR resonance for the bulge and additional higher resonances have been found to trap a negligible fraction of bulge orbits, $\ltorder 1\%$. This happens typically at very late times of the runs. The reason is that these bulge orbits lie deep inside the total gravitational potential well.  

\section{Discussion}
\label{sec:discuss} 

We run a suite of numerical simulations of stellar disks with bulges immersed in DM halos. A range of bulge masses of 10\% - 25\% of the total disk $+$ bulge mass and various degrees of initial rotation of the bulge and DM halos have been used in our models. 

Our results are as follows:

\begin{itemize}

\item In nonspinning DM halos, i.e., $\uplambda = 0$, we find a very small decrease of the maximal amplitude of stellar bars before the vertical buckling. The decrease in bar strength during the buckling, $\Delta A_2$, however, becomes smaller with increasing bulge mass, and so is the buckling amplitude, $A_{1z}$, independently of the initial bulge spin. All modeled bars have restarted their growth after buckling and reached about the same amplitude of $A_2$ after 10\,Gyr. 

\item In contrast, in spinning DM halos with $\uplambda = 0.09$, all models with the bulge mass fractions of $10-25\%$ have reduced the pre-buckling amplitude $A_2$ from $\sim 0.3$ to $\sim 0.2$. The buckling amplitude, $A_{1z}$, has been reduced dramatically in all these models by a factor of $\sim 2$ compared to the bulgeless model. Moreover, some models do not show any buckling of stellar bars, in agreement with \citet{sell20}, although our bulge parameters differ. Furthermore, stellar bar amplitude does not recover after buckling in spinning halos, in agreement with previous results \citep{long14,collier18,li23b}

\item Stellar bar instability is delayed both in more massive bulges and even more in faster spinning bulges, i.e., along the bulge mass sequence B0\,$\rightarrow $\,B25, and along the bulge spin sequence S50\,$\rightarrow $\,S95.  The stellar bars form with higher pattern speeds in higher mass bulges and even faster in bulges with a higher spin. We reiterate, as noted above, the formation of stellar bars is delayed for more massive and faster spinning bulges. This behavior is irrespective of the parent DM halo spin.

\item In order to verify that indeed the buckling of stellar bars has been suppressed in some models with spinning halos, we have used a modified method introduced by \citet{smirnov18}, and discussed here in section\,\ref{sec:alternative_buck}. This method has been found to be much more sensitive to the buckling process. Moreover, it clearly demonstrated that the buckling amplitude in spinning halos, $A_{\rm asym}$, although reduced by a factor of $\sim 2$ in bulges within the analyzed mass fraction, has a prolonged duration, as confirmed by a 2-D Laplace plane distortion map. In fact, stellar bars in these models have never fully restored their vertical symmetry within the run time.  We show that this behavior emphasizes the different morphology above and below the bar midplane. Note also that the most massive bulge model B25 in $\uplambda=0.09$ halo has never buckled, because the timescale of the bar instability is $\sim 10$\,Gyr, irrespective of the bulge spin.

\begin{figure*}
    \includegraphics[width=0.8\textwidth]{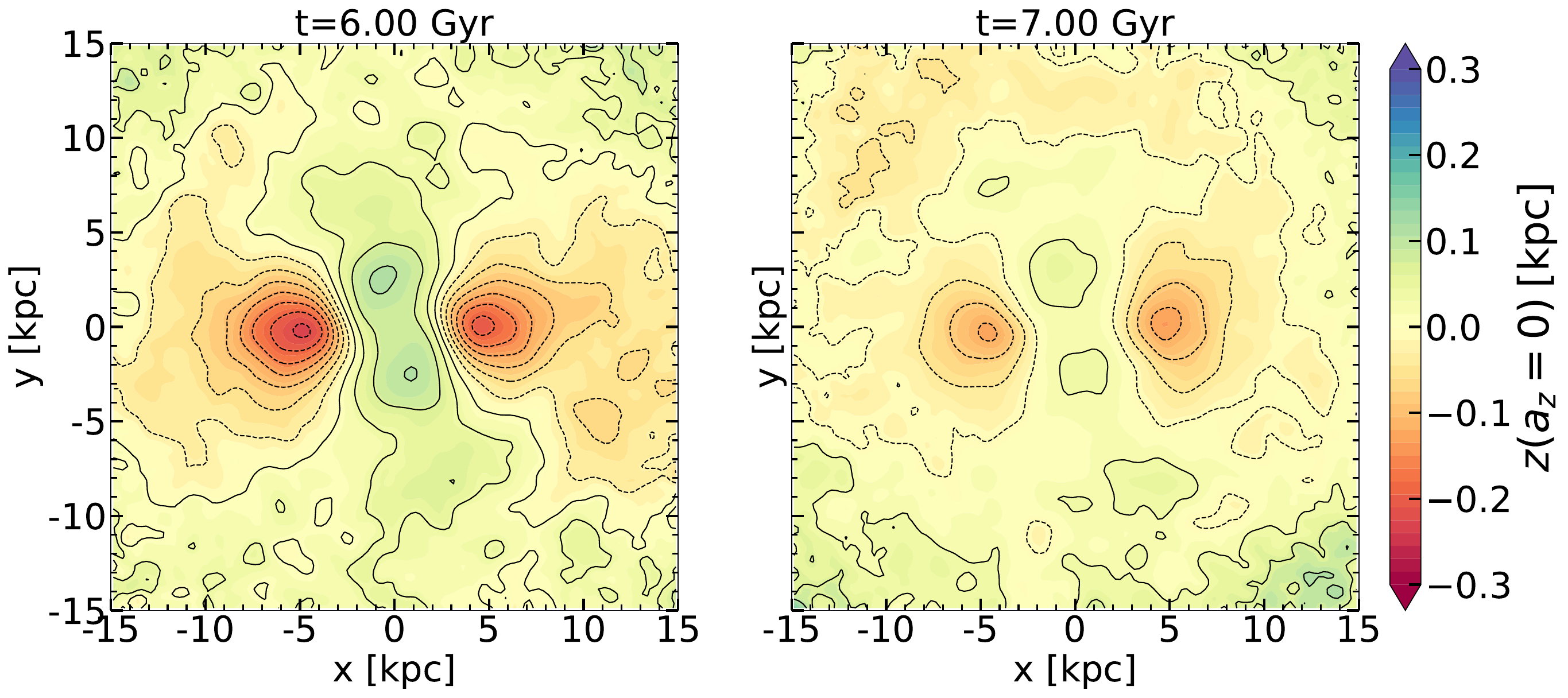}
    \caption{Same plot as Figure\,\ref{fig:az0_B10P00_1} for the model B10-S50 in $\uplambda=0.09$ halo, at the time of the maximal buckling (left) and 1\,Gyr later (right). Note that the buckling footprint is still visible as a double peak along the stellar bar (horizontal).}
    \label{fig:az0_B10P90}
\end{figure*}

\item  The angular momentum flow in the bulge-disk-halo systems is heavily dominated by the barred disk -- DM halo interactions. The lion share of angular momentum flow is from the barred disk to DM halo, as is well known from previous investigations. In this triple system, the bulges play a secondary role in $J$-transfer. Yet, the initially nonrotating bulges, S50, absorb the angular momentum from the barred disk, the intermediate rotating bulges, S75, do not change their spin, while the fast spinning bulges, S95, slowdown their initial spin. This conclusion applies both for $\uplambda = 0$ and $\uplambda = 0.09$ halos equally.

\item By applying the orbital spectral analysis for the bulge orbits, we find two competing resonances, at $\nu = 0.5$ and $\nu = -0.5$ trapping the bulge orbits, where  $\nu = (\Omega_{\mathrm{bar}} - \Omega) / \kappa$ is the dimensionless frequency ratio. The former resonance is the inner Lindblad resonance (ILR) and is the dominant one. The latter resonance involves retrograde bulge orbits only (with respect to the disk and bulge rotations), elongated perpendicular to the stellar bar. In this respect they mimic the outer Lindblad resonance (OLR) orbits, but they circulate in the opposing direction. The efficiency of this latter resonance in trapping the bulge orbits and, therefore, in transferring the angular momentum, varies with time significantly. Furthermore, both trapped prograde (at ILR) and retrograde  (at OLR) bulge orbits have been found to gain angular momentum during the evolution. In S50 models, most of the angular momentum transfer to/from the bulge proceeds via these trapped prograde and retrograde bulge orbits. Their angular momentum becomes more positive or negative, respectively.

\item In S95 models, orbits trapped at the ILR gain angular momentum. Orbits trapped at the OLR, are retrograde and gain negative angular momentum. Moreover, orbits trapped in the range of $-0.5< \nu<0$, gain negative angular momentum. While orbits in the range of $0<\nu<0.5$ lose positive angular momentum. In balance, for S95 models, the angular momentum is lost.

\end{itemize}

We now return to the focal aspect of this work. Namely, how does the bulge affect the stellar bar evolution and how does the bar reciprocate. For this purpose, we constructed sequences of the $B/T$ mass ratios and the initial bulge spin. Additional important parameters is the spin of the parent DM halo. Previous works have studied many aspects of this bulge-disk-halo system. Our main contribution in this work is to quantify the effect of the spin in both spheroidal components, understanding the flow of angular momentum between these components, and especially the role of the orbital resonances in the interaction between them.  

An interesting outcome of our analysis of these simulations is the application of the modified method of \citet{smirnov18} to quantify the vertical asymmetry of stellar bars. Not only it provides a more sensitive way to detect the vertical buckling of bars, but it reveals the long-term effect of the buckling, which has been found to extend for additional few Gyr in spinning halos. This result is supported by calculation of the distorted Laplace plane. Such a response of a barred disk was not observed before in numerical simulations and brings forward the question of what is the underlying mechanism of this prolonged violation of the vertical symmetry.

To demonstrate the prolonged buckling process in spinning halos, we refer to the face-on Laplace plane in B10-S50 model immersed in $\uplambda = 0.09$ Dm halo (Fig.\,\ref{fig:az0_B10P90}). Note that 1\,Gyr after the buckling, the double peak is still visible at slightly reduced amplitude. 

The first application of the orbital dynamics to the formation of peanut/boxy shapes of galactic bulges has determined the population of the so-called BAN/ABAN orbits \citep{pfen91}. These are 3-D orbits and their origin lies in the action of the combined horizontal ILR and vertical ILR (vILR). These orbits are 2:1 orbits (two vertical oscillations per one turn in the plane) and bifurcate vertically from the $\mathrm{x_1}$ family, and in projection on the $x-y$-plane they look like the $\mathrm{x_1}$ orbits. Particles populating these orbits are involved in the vertical oscillations which take them away from the disk midplane. 

Due to the vILR occurring at some point of the $\mathrm{x_1}$ orbit family, a vertical orbit family fork-bifurcation forms, and the possibility of a transverse break of symmetry occurs. Triggered by noise, $z$-symmetry may be broken, one side of the BAN/ABAN family pair is then more populated than the other, and a $z$-asymmetry grows. This vertical instability grows until saturation in the non-linear regime.  Under these conditions, the Laplace plane, which is determined by the potential minimum along the $z$-axis or, alternatively, by the zero-acceleration along this axis, is bent away from the disk midplane. 

The behavior of this population of orbits is closely related to the appearance of KAM surfaces\footnote{see Kolmogorov-Arnold-Moser theorem \citep[e.g.,][]{licht92}} which partly confine the chaotic motion. In 2-D systems (i.e., systems with two degrees of freedom)  these KAM surfaces isolate the phase space volumes. But they cannot do this perfectly in non-integrable systems with more than two degrees of freedom, due to the Arnold's diffusion \citep[e.g.,][]{licht92}. When close to the local integrability, the Arnold's diffusion can be slow.   

Both BAN and ABAN orbits populate different regions in phase space. When the vertical asymmetry of the potential exists, the populations of these orbits initially differ, but slowly diffuse through chaotic phase space by Arnold's diffusion until they become equal. This process can support the vertical asymmetry for prolonged time periods as discussed above (see also Figures\,\ref{fig:ver-asym} and \ref{fig:faceon-asym-b0}). 

Next, we focus on the corollaries of a bulge presence in disk galaxies. We find that whether the bulge is a recipient or donor of angular momentum in the system depends on its initial spin.  Previous studies have shown that a nonrotating bulge absorbs the angular momentum from the barred disk via gravitational torques \citep[e.g.,][]{saha12,saha13,saha16}. Such bulge correspond to our S50 models. Figure\,\ref{fig:Jz-all} confirms this evolution. Moreover, it shows that more massive bulges absorb more angular momentum. 

However, our S75 models, with an intermediate rotating bulges, show that they both emit and absorb the angular momentum (Figures\,\ref{fig:Jz-all} and \ref{fig:Jdot}), either in the DM halos with $\uplambda = 0$ or 0.09. S95 models, with fast spinning bulges, actually slow down by losing their angular momentum.

For the range of $B/T$ in our models and irrespective of the initial spin of the bulge, the amount of angular momentum absorbed or emitted by the bulge is small compared to the angular momentum exchange between the barred disk and the DM halo. This conclusion is expected as the maximal amount of angular momentum that can be stored in the bulge of fixed shape is negligible compared to the disk or DM halo (see Tables\,\ref{tab:Jmax} and \ref{tab:Jcomp}, section\,\ref{sec:Jmax} and the Appendix). Hence, the angular momentum stored in the bulge is not expected to directly affect the stellar bar evolution. However, the bulge can affect the disk evolution by other means, e.g., by destroying the harmonic motion in the core which plays an important role in the buckling process. Furthermore, the bulge can provide an initial perturbation since its dynamical timescale is the shortest in the system. Under these conditions, the development of a $m=2$ mode there can affect the development of this mode in the larger disk. 

The presence of a bulge affects the pattern speed of a bar, as was investigated by \citet{kataria19}, concluding that the slowdown is faster for more massive bulges. We confirm this result, but have noticed that the bars form also with higher $\Omega_{\rm bar}$ in the presence of more massive bulges (e.g., Figure\,\ref{fig:OmegaBar}). The net slowdown of bars is such that the $\Omega_{\rm bar}$ for a spectrum of bulge masses never intersect. So, bars in models with higher bulge masses will end up with a higher  $\Omega_{\rm bar}$ even after a 10\,Gyr run. Note also that the bar pattern speeds in spinning halos become flat quickly, stopping their decay, as expected \citep{collier18}. 

Figure\,\ref{fig:bulgeA2} displays the evolution of $A_2$ for S50, S75 and S95 bulge models. When compared with the stellar bar amplitudes in Figure\,\ref{fig:diskA2}, we observe a one-to-one correspondence. Yet, for S50 models, the bulge $m=2$ amplitude is smaller by a factor of 4 than the bar $m=2$ amplitude. For S75 models, this difference is only a factor of 2. And for S95 models, there is almost no difference between the $m=2$ mode of the bulge and the parent stellar bar for parent DM halos with $\uplambda = 0$. 

For spinning halos with $\uplambda = 0.09$, already S75 bulge models have similar amplitudes of $m=2$ modes with the bar. And for S95 models, the bulge $A_2$ is stronger compared to the associated bar by about 50\%.

The spin and mass sequences of bulges have clear effects on the bulge shapes, both in the $x-y$ and vertical planes. In the $x-y$ plane, the bulge shape measures its strength via the amplitude of the $m=2$ mode. This shape can be alternatively measured using ellipticity (see section\,\ref{sec:bshape} for definitions). 

Figure\,\ref{fig:eccentr} displays the bulge ellipticity evolution in the vertical plane. In both nonspinning and spinning halos, the bulge becomes more eccentric with increasing initial spin reaching $\epsilon\sim 0.3$ for S95 models. This appears to be the maximal ellipticity reached by a bulge nearly maximally spinning, corresponding to the axial ratio of $c/a\sim 0.7$. Such axial ratios are observed in maximally flattened bulges and elliptical galaxies \citep[e.g.,][]{binn98}.

\section{Conclusions}
\label{sec:conclusions}

We have studied the evolution of barred disks immersed in spheroidal components which include DM halos and stellar bulges. The initial conditions include a sequence of bulge masses, as well as sequences of bulge and halo spins. The analyzed $B/T\sim 0-25\%$ mass ratio range, corresponds to the one typically observed in disk galaxies, while the DM halo spin range, $\uplambda \sim 0-0.09$ to the most probable one based on the numerical simulations.

We find that the presence of a bulge affects evolution of stellar bars, including the characteristic timescale of the bar instability, the bar pattern speed and its decay, bar length, and its vertical buckling instability. The main effect of the bulge is the destruction of the harmonic core in the disk which affects the vertical buckling instability in the bar, as shown by \citet{li23a}. Moreover, the bulge does not preclude the secondary buckling of stellar bars, as expected, as it happens away from the central regions dominated by the bulge \citep{marti06}. However, as a reservoir or sink of angular momentum, the bulge plays a minor role, and the angular momentum exchange proceeds mainly between the DM halo and the barred disk. 

In the modeled range of $B/T\sim 0-25\%$ in nonspinning halos, we do not find any damping of the amplitude of the bar instability, but find a progressively increasing delay in the characteristic timescale of instability. In contrast, within spinning halos with $\uplambda = 0.09$, the bar amplitude has decreased by $\sim 1/3$ compared to halos with $\uplambda = 0$.   

We confirm the previous result by \citet{kataria19} that pattern speeds of newly formed bars appear higher for more massive bulges.  Moreover, we find that, although such bars formed with higher $\Omega_{\rm bar}$  and experiencing a slowdown, will always end up with higher pattern speeds even after a 10\,Gyr. But we do not find such a sensitivity of the stellar bar amplitude to the presence of the bulge as claimed by \citet{saha13} and \citet{sell20}.

We have applied a modified version of the parameter proposed by \citet{smirnov18} measuring the vertical buckling amplitude, $A_{\rm asym}$, and found it to be substantially more sensitive to buckling compared to the Fourier $A_{1z}$ parameter used in the literature. Using $A_{\rm asym}$, we find that the buckling amplitude has been decreased by a factor of $\sim 2$ in halos with $\uplambda = 0.09$ compared to halos with $\uplambda = 0$.

Most interestingly, we find that the buckling process triggers different responses above and below the disk midplane. This difference can be diluted by progressively more massive bulges.
In spinning halos, the buckling distortion has a prolonged amplitude tail which extends by a few Gyr. We have tested this phenomenon and verified it by using the distortion of the Laplace plane. We relate this prolonged asymmetry with the evolution of the population of the 3-D BAN/ABAN orbits and slow Arnold's diffusion in our systems. This prolonged asymmetry can potentially have observational corollaries.

The angular momentum flow in the barred disk-bulge-halo systems proceeds mainly between the barred disk and the DM halo, with bulge in the mass range of $B/D\sim 0-25\%$ playing a minor role. As verified using orbital spectral analysis,  non-spinning bulges gain their spin from the stellar bar, and do it mainly via the ILR, while fast-spinning bulges lose their spin via  prograde orbits which lie between the CR and the ILR, and retrograde orbits trapped by the OLR.

\section*{Acknowledgements}

We thank Phil Hopkins for providing us with the latest version of GIZMO and Angela Collier for sharing some of the analysis software. I.S. is grateful for a generous support from the International Joint Research Promotion Program at Osaka University, and for hospitality of Kavli Institute for Theoretical Physics (KITP) during his prolonged visit there. This work has also been supported in part by the JSPS KAKENHI grant 16H02163 (to I.S.) and by the NSF under Grant PHY-1748958 to KITP. Simulations have been performed using the University of Kentucky Lipscomb Computing Cluster. We are grateful to Vikram Gazula at the Center for Computational Studies at the University of Kentucky for help with the technical issues during the LCC runs.

\section*{Data Availability}

The data presented in this work can be obtained upon reasonable request.



\bibliographystyle{mnras}
\bibliography{paper} 




\appendix

\section*{Appendix:Maximum angular momentum of galaxy components}

\renewcommand{\theequation}{A\arabic{equation}}

Here we provide tighter bounds than usual on the maximum angular momentum than a system of particles
in virial equilibirum can contain. We use the standard Cartesian position and velocity coordinates $\{x_i,y_i,z_i\}$ 
and
$\{{v_x}_i,{v_y}_i,{v_z}_i\}$ for particle $i$ of mass $m_i$, where the centers of mass and velocity are set at the origin.

\cite{sundman1913} and \cite{muller1986} showed that in any steady collection of material particles, where 
the extent of its mass distribution is expressed  by its moment of inertia,
\begin{equation}
I = \sum_i m_i (x_i^2+y_i^2+z_i^2),
\end{equation}
and its total kinetic energy,
\begin{equation}
   T = \frac{1}{2}\sum_i m_i ({v_x}_i^2+{v_y}_i^2+{v_z}_i^2), 
\end{equation} 
the total angular momentum squared is bound by the product of the latter quantities, 
\begin{equation}
    J^2 \leq 2 T I  =\sum_i m_i ({v_x}_i^2+{v_y}_i^2+{v_z}_i^2) \sum_i m_i (x_i^2+y_i^2+z_i^2).
\end{equation}

If we know the orientation of the spin vector, say along the $z$-axis, we can remove 
the irrelevant coordinates $z_i$ and ${v_z}_i$ and derive tighter inequalities as shown by 
\citet{pfen19}.  The first inequality, named $P_1$ is 
\begin{equation}
    J^2 \leq  \sum_i m_i (x_i^2+y_i^2) \sum_i m_i({v_x}_i^2+{v_y}_i^2) \equiv P_1^2.
    \label{eq:P1}
\end{equation}
The first sum is the moment of inertia orthogonal to the $z$-axis, the second sum twice the horizontal kinetic energy. 

A tighter inequality, named $P_2$, can be formulated as,
\begin{equation}
    J^2 \leq  \sum_i m_i (x_i^2+y_i^2) \sum_i m_i \frac{(x_i{v_y}_i - y_i {v_x}_i)^2}{x_i^2+y_i^2} \equiv P_2^2.
    \label{eq:P2}
\end{equation}
The second sum here is twice the rotational kinetic energy around the $z$-axis, which is smaller or equal to 
the total horizontal kinetic energy in $P_1$, so $P_2 \leq P_1$.

For the reason that $J^2$ is invariant by a swap of the velocity and position coordinates, 
a distinct inequality can be derived from $P_2$, inequality $P_3$, 
\begin{equation}
    J^2 \leq  \sum_i m_i({v_x}_i^2+{v_y}_i^2)\sum_i m_i \frac{(x_i{v_y}_i - y_i {v_x}_i)^2}{{v_x}_i^2+{v_y}_i^2}  \equiv P_3^2 .
    \label{eq:P3}
\end{equation}
Since $P_1$ is swap symmetric and $P_2 \leq P_1$, then by symmetry $P_3 \leq P_1$. 
However, depending on configuration each of $P_2$ or $P_3$ may be different, therefore
one of the two is the smallest.  
Using the minimum of $P_2$ and $P_3$, 
\begin{equation}
    |J| \leq \frac{1}{2}\left(P_2+P_3-|P_2-P_3|\right)  \equiv P_{23} , 
    \label{eq:P23}
\end{equation}
we obtain the tightest bound on the maximum angular momentum 
for each of the bulge, disk and halo galaxy components, expressed with components 
of their respective moment of inertia and kinetic energy tensors.





\bsp	
\label{lastpage}
\end{document}